\documentclass{article}

\usepackage{microtype}
\usepackage{graphicx}
\usepackage{subfigure}
\usepackage{booktabs} 

\usepackage{hyperref}

\usepackage[accepted]{icml2025}

\usepackage{amsmath}
\usepackage{amssymb}
\usepackage{mathtools}
\usepackage{amsthm}

\usepackage{amsfonts}
\usepackage{amsmath}
\usepackage{amssymb}
\usepackage{MnSymbol}
\usepackage{wasysym}
\newcommand{\independent}{\perp\!\!\!\perp}

\usepackage[capitalize,noabbrev]{cleveref}

\theoremstyle{plain}
\newtheorem{theorem}{Theorem}[section]
\newtheorem{proposition}[theorem]{Proposition}
\newtheorem{lemma}[theorem]{Lemma}

\theoremstyle{definition}
\newtheorem{definition}[theorem]{Definition}

\newtheorem{hypothesis}[theorem]{Hypothesis}
\theoremstyle{remark}

\usepackage[textsize=tiny]{todonotes}

\newcommand{\red}[1]{\textcolor{red}{#1}}
\usepackage{arydshln}
\icmltitlerunning{Permutation-Free High-Order Interaction Tests}

\begin{document}

\twocolumn[
\icmltitle{Permutation-Free High-Order Interaction Tests}



\icmlsetsymbol{equal}{*}



\begin{icmlauthorlist}
\icmlauthor{Zhaolu Liu}{icl}
\icmlauthor{Robert L. Peach}{rob1,rob2}
\icmlauthor{Mauricio Barahona}{icl}
\end{icmlauthorlist}

\icmlaffiliation{icl}{Department of Mathematics, Imperial College London, United Kingdom}
\icmlaffiliation{rob1}{Department of Neurology, University Hospital Würzburg, Germany}
\icmlaffiliation{rob2}{Department of Brain Sciences, Imperial College London, United Kingdom}

\icmlcorrespondingauthor{Mauricio Barahona}{m.barahona@imperial.ac.uk}

\icmlkeywords{Machine Learning, ICML}

\vskip 0.3in
]



\printAffiliationsAndNotice{}  

\begin{abstract}
Kernel-based hypothesis tests offer a flexible, non-parametric tool to detect high-order interactions in multivariate data, beyond pairwise relationships. Yet
the scalability of such tests 
is limited by the computationally demanding permutation schemes used to generate null approximations. 
Here we introduce a family of permutation-free high-order tests for joint independence and partial factorisations of $d$ variables. Our tests eliminate the need for permutation-based approximations by leveraging V-statistics and a novel cross-centring technique to yield test statistics with a standard normal limiting distribution under the null. 
We present implementations of the tests and showcase their efficacy and scalability through synthetic datasets. We also show applications inspired by causal discovery and feature selection, which highlight both the importance of high-order interactions in data and the need for efficient computational methods. 
\end{abstract}

\section{Introduction}

Many complex real-world systems involve interactions beyond pairwise relationships, necessitating an explicit understanding of interactions among three or more entities~\citep{battiston2020networks,  battiston2021physics,rosas2022disentangling, thibeault2024low}.
Such high-order interactions amongst $d>2$ variables have been shown to affect processes in
social~\citep{benson2016higher}, ecological~\citep{grilli2017higher} and biological systems~\citep{luff2024neuron, arnaudon2022connecting}, often in ways that cannot be reduced to combinations of pairwise effects~\citep{schaub2020random,bick2023higher}.
Given the relative scarcity of relational datasets that record high-order phenomena, it is important to develop unsupervised, mathematically grounded methods for their detection in observational data.
A growing body of work has approached this challenge using tools from 
information theory~\citep{rosas2020operational,rosas2024characterising}, persistent homology~\citep{santoro2023higher,santoro2024higher}, Bayesian inference~\cite{young2021hypergraph} and Hamiltonian-based functions~\citep{schneidman2003network}. 

Within the statistical and machine learning literature,
kernel-based tests provide a flexible, non-parametric framework for identifying relationships in data. The detection of pairwise independence between two variables using $\mathrm{HSIC}$~\citep{gretton2007kernel} 
has found broad applicability in tasks such as generative adversarial networks~\citep{binkowski2018demystifying}, feature selection~\citep{song2012feature}, and transfer learning~\citep{long2015learning}.
Beyond pairwise independence, kernel-based tests have been 
extended to assess joint independence among $d$ variables via $\mathrm{dHSIC}$~\citep{pfister2018kernel}, as well as Lancaster interactions among three variables~\citep{sejdinovic2013kernel}, which were later generalised to $d$-order Streitberg interactions~\citep{liu2023interaction}. However, these tests typically rely on computationally expensive permutation or bootstrap schemes (involving $100\leq p\leq 1000$ permutations), reducing their applicability to datasets with larger number of variables, where the subsets of variables to be tested for interactions explode combinatorially.

Here, we tackle this limitation on scalability by introducing a family of permutation-free high-order tests that retain the flexibility of kernel-based methods while eliminating the  computational burden of null-distribution approximation. To note, our approach applies to the general settings where one seeks to determine factorisation properties among subsets of variables (e.g., Lancaster and Streitberg interactions), rather than merely rejecting or accepting joint independence. 

\paragraph{Contributions} Our main contributions can be summarised as follows: (1) We define the permutation-free joint independence test, $\overline{\mathrm{x}}\mathrm{dHSIC}$ for any $d$. (2) To address the combinatorial challenges of testing for high-order factorisation, we introduce the cross-centring technique to develop the permutation-free $d$-order Lancaster factorisation test, $\overline{\mathrm{x}}\mathrm{LI}$. (3) Building on $\overline{\mathrm{x}}\mathrm{LI}$, we derive a permutation-free Streitberg interaction test, $\overline{\mathrm{x}}\mathrm{SI}$. (4) We show that, for the special case $d=2$, our  high-order tests have a more concise mathematical formulation than the existing $\overline{\mathrm{x}}\mathrm{HSIC}$~\citep{shekhar2023permutation}. (5) Through empirical evaluations in machine learning applications, we demonstrate that the permutation-free tests achieve substantial computational speedup, exceeding 100-fold, over their permutation-based counterparts.

\paragraph{Paper Structure} 
Section~\ref{sec:preliminaries}  introduces joint independence and partial factorisation as two different generalisations of pairwise relationships, and briefly reviews kernel-based tests. In Section~\ref{sec: permfree_highorder}, we develop our permutation-free joint independence test for any $d$, $\overline{\mathrm{x}}\mathrm{dHSIC}$, and our permutation-free factorisation tests for any $d$, $\overline{\mathrm{x}}\mathrm{LI}$ and $\overline{\mathrm{x}}\mathrm{SI}$.
We relate our proposed methods to existing work in Section~\ref{sec: related_work}, and provide numerical validations in
Section~\ref{sec: experiments}.
We conclude by discussing directions for future research in Section~\ref{sec: discussion}.
\section{Preliminaries}\label{sec:preliminaries}
\subsection{Statistical Interactions}\label{sec: statistical_interactions}
Throughout this paper, we consider multivariate probability distributions of a set of $d$ variables $D = \{X^1, \ldots, X^d\}$ denoted $\mathbb{P}_{X^1, \ldots, X^d} =: \mathbb{P}_{1\cdots d}$. 

\paragraph{Pairwise Interaction} The simplest statistical interaction emerges for $d=2$. In this case, a lack of statistical interaction is equivalent to pairwise independence, which is determined by the difference between the joint distribution $\mathbb{P}_{12}$ and the product of the marginals $\mathbb{P}_{1}\mathbb{P}_2$: 
\begin{align}
\label{eqn: pairwise_independence}
\Delta^2 \mathbb{P} =   \mathbb{P}_{12}-\mathbb{P}_{1}\mathbb{P}_{2}\,.
\end{align}
Hence $\Delta^2 \mathbb{P} =0 \iff X^1 \independent X^2$ (pairwise independence).

\paragraph{Joint Independence} One way to generalise the above notion to $d>2$ variables is to focus on joint independence:
\begin{align}\label{eqn: joint_independence}
    \Delta_I^d \mathbb{P}=\mathbb{P}_{1\cdots d} -\prod_{i=1}^d \mathbb{P}_{i}\,.
\end{align}
Clearly, $\Delta_I^d \mathbb{P}$ vanishes if and only if the $d$ variables are jointly independent, a property that has been employed in Independent Component Analysis~\citep{hyvarinen2000independent}, causal discovery~\citep{pfister2018kernel,laumann2023kernel},  and variational autoencoders~\citep{lopez2018information}.
However, for $d\geq 3$, joint independence is uninformative as a criterion for the presence of high-order interactions. 
For example, three of the four factorisations of $\mathbb{P}_{123}$ ($\mathbb{P}_{1}\mathbb{P}_{23}$, $\mathbb{P}_{2}\mathbb{P}_{13}$ and $\mathbb{P}_{3}\mathbb{P}_{12}$) are neither full 3-way interactions nor jointly independent like $\mathbb{P}_{1}\mathbb{P}_{2}\mathbb{P}_{3}$.

\paragraph{Lancaster Interaction} An alternative generalisation interprets the vanishing condition of $\Delta^2 \mathbb{P}$ as a factorisation of the joint distribution into disjoint components. To capture this perspective, 
the Lancaster interaction for $d=3$ variables is defined as:
\begin{equation}\label{eqn: lancaster_3}
    \Delta_L^3 \mathbb{P} = \mathbb{P}_{123}-\mathbb{P}_{1}\mathbb{P}_{23}-\mathbb{P}_{2}\mathbb{P}_{13} - \mathbb{P}_{3}\mathbb{P}_{12}+2\mathbb{P}_{1}\mathbb{P}_{2}\mathbb{P}_{3}\,.
\end{equation}
If $\mathbb{P}_{123}$ can be factorised into \textit{any} of $\mathbb{P}_{1}\mathbb{P}_{23}, \mathbb{P}_{2}\mathbb{P}_{13}, \mathbb{P}_{3}\mathbb{P}_{12}$ or $\mathbb{P}_{1}\mathbb{P}_{2}\mathbb{P}_{3}$, then $\Delta_L^3 \mathbb{P}=0$ (see Appendix~\ref{app: lancaster_3}). \citet{Lancaster} rewrote and generalised this measure for $d$-order interactions as
\begin{equation}\label{eqn: lancaster_interaction}
    \Delta_L^d \mathbb{P}=\prod_{i=1}^d\left(\mathbb{P}_{i}^*-\mathbb{P}_{i}\right)\,, 
\end{equation}
where the $\mathbb{P}_{i}^*$ are defined implicitly by $\mathbb{P}_{i}^*\mathbb{P}_{j}^*\cdots \mathbb{P}_{k}^* = \mathbb{P}_{ij\cdots k}$ and
$\mathbb{P}_{i}^*\mathbb{P}_{j}\cdots \mathbb{P}_{k} = \mathbb{P}_{i} \mathbb{P}_{j}\cdots \mathbb{P}_{k}$.

However, for $d\geq 4$, $\Delta_L^d \mathbb{P}$ fails to capture certain factorisations, e.g.,  if $\mathbb{P}_{1234}=\mathbb{P}_{12}\mathbb{P}_{34}$ then 
$\Delta_L^d \mathbb{P}=(\mathbb{P}_{12}-\mathbb{P}_1\mathbb{P}_2)(\mathbb{P}_{34}-\mathbb{P}_3\mathbb{P}_4)$ which is not zero in general~\citep{streitberg1990lancaster}. 
Despite this limitation, \citet{liu2023interaction} proved that for $d \geq 4$, $\Delta_L^d \mathbb{P}=0$ only if the factorisation of $\mathbb{P}_{1\cdots d}$ includes singleton variables, highlighting its partial informativeness in capturing structural dependencies.

\begin{figure}[t]
  \centering
  \includegraphics[width=0.95\linewidth]{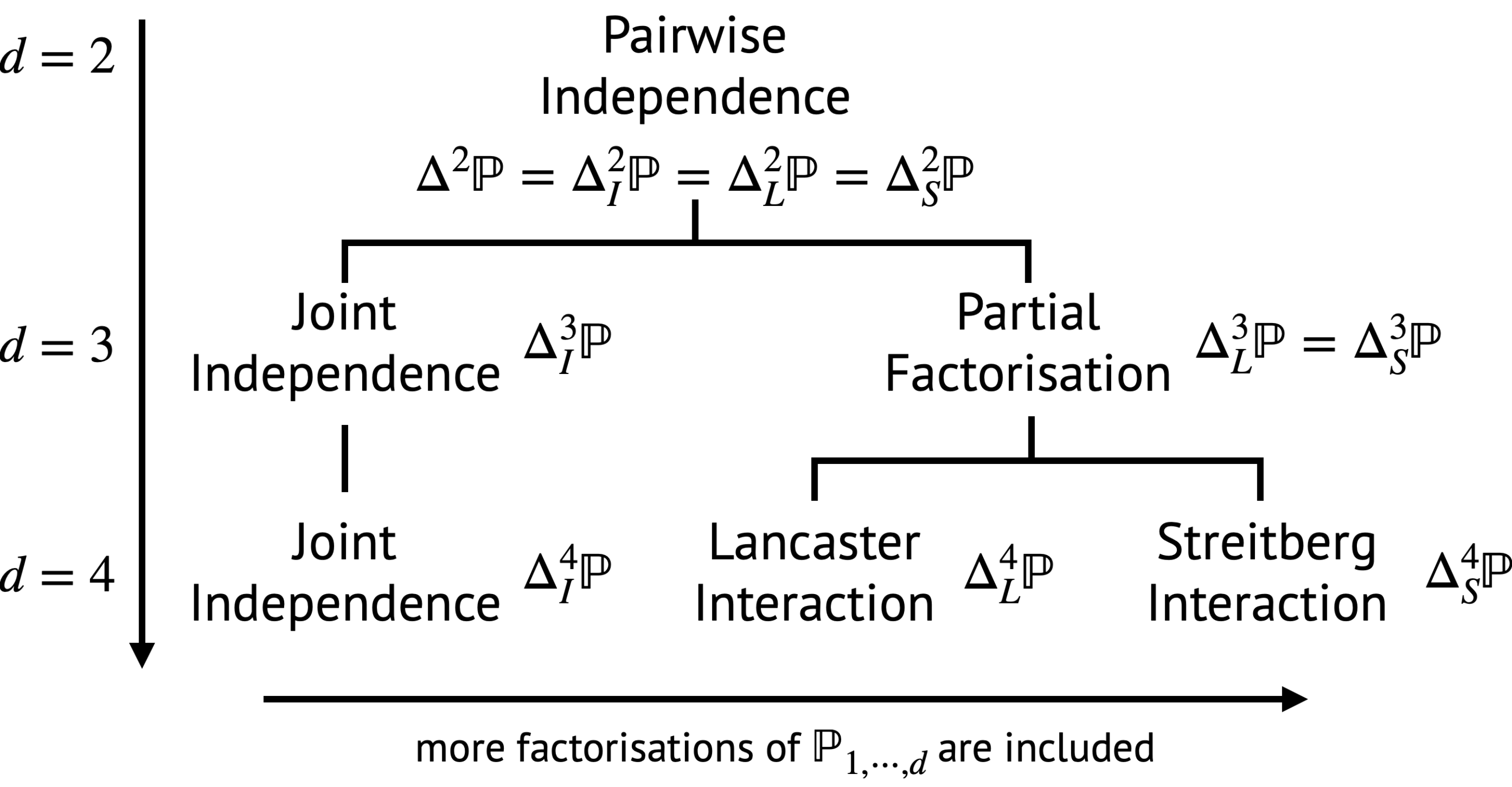}
  \caption{\textbf{Generalisations of pairwise dependence to high-order interactions.} All three measures of statistical interaction (joint independence, Lancaster, Streitberg) are the same for $d=2$. Joint independence captures less information for $d\geq 3$. 
  Only Streitberg interaction considers all factorisations for $d\geq 4$.}\label{fig: family_tree}
\end{figure}

\paragraph{Streitberg Interaction}
To account for \emph{all} factorisations of $\mathbb{P}_{1\cdots d}$ , the Streitberg interaction was introduced \citep{streitberg1990lancaster}. Let $\pi = b_1\vert b_2\vert\ldots\vert b_{|\pi|}$ denote a partition of the set $D$ of $d$ variables into $|\pi|$ blocks $b_j$, and let $\mathbb{P}_{\pi} := \prod_{j=1}^{|\pi|} \mathbb{P}_{b_j}$. The Streitberg interaction for $d$ variables is: 
\begin{equation}\label{eqn: streitberg_interaction}
\Delta_S^d \mathbb{P} = \sum_{\mathclap{\pi \in \Pi(D)}} \left(|\pi| - 1\right)! \, (-1)^{|\pi| - 1} \mathbb{P}_{\pi}\,,
\end{equation}
where the sum extends over $\Pi(D)$, the set of all partitions of the set $D$.
Crucially, $\Delta_S^d \mathbb{P} = 0$ only if $\mathbb{P}_{1\cdots d}$ can be factorised in any way~\citep{streitberg1990lancaster}.

Figure~\ref{fig: family_tree} summarises these three perspectives of high-order interactions, showing how they all coincide for $d=2$; begin to diverge for $d=3$; and  Streitberg interactions are unique in capturing all possible factorisations for $d \geq 4$. For details on derivations via lattice theory, see~\citet{liu2024information}.

\subsection{Kernel-Based Tests}\label{sec: pairwise_tests}
To transform the interaction measures from Section~\ref{sec: statistical_interactions} into non-parametric test statistics, we embed the underlying distributions into reproducing kernel Hilbert spaces (RKHS). 
Recall that a symmetric, positive-definite kernel $k^i: \mathcal{X}^i \times \mathcal{X}^i \rightarrow \mathbb{R}$ induces an associated RKHS $\mathcal{H}^i$  with the reproducing property. For any $X^i\in \mathcal{X}^i$, let 
$\phi^i(X^i) = k^i(X^i,\cdot)$ denote the canonical feature map. The kernel mean embedding of a distribution  
$\mathbb{P}_{i}$
is then given by $\mu_{\mathbb{P}_{i}} = \mathbb{E}_{X^i}[\phi^i(X^i)]$. This embedding satisfies $\mathbb{E}_{X^i} f({X^i})=\left\langle f, \mu_{\mathbb{P}_i}\right\rangle_\text{HS}$ for all $f$ in  $\mathcal{H}^i$, where $\left\langle\cdot,\cdot\right\rangle_\text{HS}$ is the Hilbert-Schmidt inner product. If $k^i$ is \emph{characteristic}, the map $\mu_{\mathbb{P}_{i}}$ is injective, ensuring that the norm of the signed measure is zero if and only if the measure itself is zero~\citep{gretton2007kernel, Muandet2017kernelreview}.

Using such embeddings, all high-order interaction measures in Section~\ref{sec: statistical_interactions} can be naturally converted into kernel-based test statistics by constructing kernel mean embeddings of the relevant distributions, and computing their Hilbert-Schmidt norms. Adopting this methodology, the measures in Eq.~\eqref{eqn: joint_independence}, Eq.~\eqref{eqn: lancaster_interaction} and Eq.~\eqref{eqn: streitberg_interaction} have been developed into, respectively, the $\mathrm{dHSIC}$~\citep{pfister2018kernel}, the Lancaster interaction test ($\mathrm{LI}$)~\citep{sejdinovic2013kernel, liu2023interaction}, and the Streitberg interaction test ($\mathrm{SI}$)~\citep{liu2023interaction}.

A key practical issue is the fact that all of these test statistics are degenerate under their respective null hypotheses. To obtain valid rejections, one typically employs a permutation procedure to approximate the null distribution, or uses heuristic strategies such as a Gamma approximation (as in $\mathrm{HSIC}$~\citep{gretton2007kernel} and $\mathrm{dHSIC}$~\citep{pfister2018kernel}). 
Permutation-based approaches involve $p$ runs (typically $p>100$) to achieve an accurate approximation of the null distribution, and are thus computationally expensive for large sample sizes or multiple variables. This computational burden underscores the need for alternative methods that are both theoretically principled and computationally efficient.

Recently, \citet{shekhar2023permutation} introduced $\overline{\mathrm{x}}\mathrm{HSIC}$, a variant of HSIC for $d=2$ that avoids the need for permutations by establishing that its null distribution converges to a standard normal.
Inspired by this approach, we develop a family of permutation-free high-order tests in Section~\ref{sec: permfree_highorder}.

\section{Permutation-Free High-Order Tests}\label{sec: permfree_highorder}
\subsection{Joint Independence}
To extend the permutation-free pairwise independence measure of~\citet{shekhar2023permutation}
to the multivariate setting, 
we first define the unnormalised permutation-free $\mathrm{dHSIC}$ using the data-splitting technique:
\begin{align*}
    \mathrm{xdHSIC} = \left\langle \hat{\mu}^1_{\mathbb{P}_{1\cdots d}} - \hat{\mu}^1_{\mathbb{P}_1\cdots\mathbb{P}_d}\,,\,\hat{\mu}^2_{\mathbb{P}_{1\cdots d}} - \hat{\mu}^2_{\mathbb{P}_1\cdots\mathbb{P}_d}\right\rangle_{\mathrm{HS}},
\end{align*}
where $\hat{\mu}^1_{(\cdot)}$ and $\hat{\mu}^2_{(\cdot)}$ are the empirical embeddings in $\mathrm{dHSIC}$, estimated from two disjoint sample sets:
the first half $\mathcal{S}^d_1=\left\{\left(x^1_i,\cdots, x^d_i\right): 1\leq i \leq n \right\}$ and the second half  $\mathcal{S}^d_2=\left\{\left(x^1_i,\cdots, x^d_i\right): n+1\leq i \leq 2n \right\}$ of the $i.i.d.$ samples.

Unlike $\overline{\mathrm{x}}\mathrm{HSIC}$  in \citep{shekhar2023permutation}, which was based on U-statistics, here we adopt V-statistics to reduce computational overhead, while retaining robust estimation accuracy~\citep{pfister2018kernel}. 
Specifically, we form the $2n \times 2n$ kernel matrices $K^i$ with entries $K^i_{ab}=k^i(x^i_a, x_b^i)$ for $1\leq a, b\leq 2n$ where $\{x^i_a\}^{2n}_{a=1}$ are $i.i.d.$ samples for the $i$-th variable. In our construction, the $n \times n$ submatrices on the diagonal  of $K^i$ (with entries $K^i_{ab}$ where both samples  $a,b$ come from either $\mathcal{S}^d_1$ or from $\mathcal{S}^d_2$) do not contribute to the estimate. Instead, we only use the $n \times n$ upper-right off-diagonal block (lower-left block is redundant due to symmetry), which measures distances between samples from the two disjoint subsets.
We then define $\overline{\mathrm{x}}\mathrm{dHSIC}$, the normalised permutation-free 
$\mathrm{dHSIC}$.
\begin{definition}
     Let $\mathcal{K}^j := K^j_{[1:n, n+1:2n]}$ be the off-diagonal $n\times n$ submatrix of $K^j$. 
     Then $\overline{\mathrm{x}}\mathrm{dHSIC}$ is defined as:
    \begin{align*}
    &\overline{\mathrm{x}}\mathrm{dHSIC}=\frac{\sqrt{n}\cdot\mathrm{xdHSIC}}{s^2_I}, \quad\text{with}\\
    &\mathrm{xdHSIC} = \frac{1}{n^2} \mathbf{1}^T
    \left(\bigcircle_{j=1}^d \mathcal{K}^j\right)\mathbf{1} 
    +\frac{1}{n^{2d}}\left(\bigcircle_{j=1}^d \mathbf{1}^T \mathcal{K}^j\mathbf{1}\right)\\
    &\qquad\qquad - \frac{1}{n^{d+1}}\mathbf{1}^T \left(\bigcircle_{j=1}^d \mathcal{K}^j\mathbf{1}\right) -\frac{1}{n^{d+1}} \left(\bigcircle_{j=1}^d \mathbf{1}^T \mathcal{K}^j\right)\mathbf{1}\,,
    \\
    &s_I^2=\frac{1}{n} \bigg\lVert\frac{1}{n}\left(\bigcircle_{j=1}^d \mathcal{K}^j\right)\mathbf{1}  
    +\frac{1}{n^{2d-1}}\mathcal{K}^1 \mathbf{1}\left(\bigcircle_{j=2}^d \mathbf{1}^T \mathcal{K}^j\mathbf{1}\right)\\
    &\qquad\quad- \frac{1}{n^{d}} \left(\bigcircle_{j=1}^d \mathcal{K}^j\mathbf{1}\right)- 
    \frac{1}{n^{d}}\left(\bigcircle_{j=1}^d {\mathcal{K}^j}^T\mathbf{1}\right)
    -\mathrm{xdHSIC}\,\mathbf{1}\bigg\rVert^2,
\end{align*}
\end{definition}
where the operator $\bigcircle$ represents the Hadamard product and $\textbf{1}$ is the $n\times 1$ vector of ones.

$\overline{\mathrm{x}}\mathrm{dHSIC}$ can be used to test $d$-order joint independence:
\begin{hypothesis} (Joint Independence)
    \begin{align*}
    &H_0:  \mathbb{P}_{1,\cdots,d} = \Pi_{i=1}^d \mathbb{P}_i\\
    &H_1:  \mathbb{P}_{1,\cdots,d} \neq \Pi_{i=1}^d \mathbb{P}_i \, .
\end{align*}
\end{hypothesis}
Then we have: $H_0 \iff \Delta_I^d P = 0 \iff \overline{\mathrm{x}}\mathrm{dHSIC} = 0.$

The construction of $\overline{\mathrm{x}}\mathrm{dHSIC}$ ensures it follows $\mathcal{N}(0,1)$ under the null~\citep{shekhar2023permutation}, thus enabling the test with a single computation in contrast with 
$p$ permutations for standard $\mathrm{dHSIC}$. Importantly, its time complexity matches that of a single $\mathrm{dHSIC}$ computation.
\begin{proposition}\label{prop: xdHSIC_time}
$\overline{\mathrm{x}}\mathrm{dHSIC}$ can be computed in $\mathcal{O}(dn^2)$.
\end{proposition}
To prove this, it is sufficient to show that both the numerator, $\mathrm{xdHSIC}$, and the denominator, $s^2_I$, can be computed in $\mathcal{O}(dn^2)$,
which follows from an argument in \citet{pfister2018kernel} (see Appendix~\ref{app: proof}). 

\subsection{Partial Factorisation}
A limitation of the joint independence test is that, if the null hypothesis is rejected, it does not distinguish between \emph{partial factorisation}, where disjoint subsets of variables indicate low-order interactions, and \emph{true non-factorisation}, where the joint distribution cannot be factorised in any way, implying that all variables interact.
To address this limitation, more sophisticated criteria have been developed~\citep{streitberg1990lancaster}, and the corresponding kernel-based tests have been introduced~\citep{sejdinovic2013kernel, liu2023interaction}. However, these methods encounter degenerate cases, which necessitate permutation-based resampling---a step that is eliminated in the 
permutation-free Lancaster and Streitberg interaction tests introduced below.

\textit{Centring:} Our approach relies on \textit{matrix centring}, which is widely used in both kernel-based~\citep{gretton2007kernel,sejdinovic2013kernel} and distance-based~\citep{chakraborty2019distance,bottcher2020dependence} high-order interaction tests to improve computational efficiency and simplify notation. For a $n\times n$ kernel matrix $K$, its centred version is $\tilde{K} = CKC$, where $C = I_n - \frac{1}{n}\textbf{1}\textbf{1}^T$, where $I_n$ is the $n \times n$ identity matrix. 

\textit{Equivalence:} Two kernels 
are said to be equivalent if they induce the same semimetric on the domain~\citep{sejdinovic2013equivalence}.
In particular, the centred kernel 
is translation-invariant and thus is equivalent to the original kernel.

\textbf{Cross-centring:} 
To extend the permutation-free scheme to a broader class of measures, we introduce an analogous \textit{cross-centring}, which arises naturally under data splitting.
\begin{definition}(Cross-centring)  
Let 
$\mathbf{1}_u= 
\tbinom{1}{0}
\otimes \textbf{1}$ and 
$\mathbf{1}_\ell=
\tbinom{0}{1}
\otimes \textbf{1}$, 
where $\otimes$ denotes the Kronecker product,
and define the matrices 
$C_u = I_{2n} - \frac{1}{n} \mathbf{1}_u{\mathbf{1}_u}^T$, and $C_\ell = I_{2n} - \frac{1}{n} \mathbf{1}_\ell {\mathbf{1}_\ell}^T$,  where $I_{2n}$ is the $2n\times 2n$ identity matrix. 
For a $2n\times 2n$ kernel matrix $K$ with indices $1:n$ associated with $\mathcal{S}^d_1$ and indices $n+1:2n$  associated with $\mathcal{S}^d_2$, its cross-centred  version is given by
$\overline{K} = C_u K C_{\ell}$.
\end{definition}

\begin{lemma}\label{lem: cross-centring}
    The cross-centred kernel 
    \begin{align*}
        \overline{k}(x, x') = \left\langle \phi(x) - \frac{1}{n}\sum_{i=1}^n \phi (x_i),\, \phi(x') - \frac{1}{n}\smashoperator[r]{\sum_{j=n+1}^{2n}} \phi (x_j')\right\rangle, 
    \end{align*}
    where $\phi(\cdot)$ is the canonical feature map, is equivalent to $k(x, x')$ as $n\rightarrow\infty$.
    See Appendix~\ref{app: proof}.
\end{lemma}
Having established this asymptotic equivalence,
we express our test statistics in terms of cross-centred kernel matrices.

\subsubsection{Lancaster Interaction Test}
We now derive the $d$-order permutation-free Lancaster interaction test. The Lancaster criterion can be used to detect factorisations involving at least one singleton variable~\cite{liu2023interaction}. To formalise the test, we introduce the following null hypothesis:
\begin{hypothesis}(Lancaster Factorisation)
    \begin{align*}
    &H_0: \bigvee_{\pi_L} H_{\pi_L} &\forall\,|\pi_L| = 2 \\
    &H_1: \bigwedge_{{\pi_L }} \neg H_{\pi_L} &\forall\,|\pi_L| = 2
    \end{align*}
where $H_{\pi_L}$ is the event such that the joint distribution factorises according to a partition $\pi_L$ containing exactly one singleton,
whereas
$\neg H_{\pi_L}$ indicates that such a factorisation does not hold. If any of the $d$ subhypotheses in $H_0$ is true, then $\Delta^d_L \mathbb{P} = 0$. Hence if all $d$ subhypotheses are rejected, a $d$-order Lancaster interaction is detected.
\end{hypothesis}

Note that even though $H_0$ contains only $d$ subhypotheses, which correspond to the factorisations with one singleton, all other factorisations that satisfy the Lancaster vanishing condition
are actually subsumed by them~\cite{liu2023interaction}.

We can now define the normalised permutation-free statistic associated with a subhypothesis $H_{\pi_L}$ of $H_0$.  
\begin{definition}
Let $H_{\pi_L}: \mathbb{P}_{1\cdots d}=\mathbb{P}_{m}\mathbb{P}_{1,\cdots,m-1, m+1, \cdots, d}=\mathbb{P}_{\pi_L}$ for some $1\leq m\leq d$.
Under the null of $H_{\pi_L}$, the normalised permutation-free Lancaster interaction, $\overline{\mathrm{x}}\mathrm{LI}$, is: 
    \begin{align*}
        \overline{\mathrm{x}}\mathrm{LI}&=\frac{\sqrt{n} \cdot\mathrm{xLI}}{s_L}\,,\quad
         \text{with}
        \\
        \mathrm{xLI} &= \frac{1}{n^2}\mathbf{1}_u^T\left(\overline{K}^m\circ \overline{\bigcircle_{i\in D/m} \overline{K}^i}\right)\mathbf{1}_\ell \,, \\     
        s_L^2&=\frac{1}{n^3} \mathbf{1}_u^T\left\lVert\left(\overline{K}^m\circ \overline{\bigcircle_{i \in D/m} \overline{K}^i}\right)\mathbf{1}_\ell\right\rVert^{\circ 2} -\mathrm{xLI}^2,
    \end{align*}
    where $D/m$ denotes all $d$ variables except $X^m$, the operators $\circ, \bigcircle$ represent Hadamard products, and the overline $\overline{(\cdot)}$ indicates cross-centring.
\end{definition}
\textit{\underline{Remark:}} In our permutation-free procedure we have a different test statistic for each subhypothesis, all with a \emph{fixed} null distribution (standardised to $\mathcal{N}(0, 1)$).

To reject the null $H_0$ for Lancaster factorisation, we carry out $d$ subtests (one for each of the singleton factorisations), each with complexity $\mathcal{O}(dn^2)$. 
\begin{proposition}\label{prop: xLI_time}
        The time complexity of $\overline{\mathrm{x}}\mathrm{LI}$ for all $H_{\pi_L}$ across any $d$ is $\mathcal{O}(dn^2)$.
\end{proposition}
The composite hypothesis $H_0$ is rejected only if all subhypotheses are individually rejected, and, as this has been proven to be less conservative than the Bonferroni correction~\cite{rubenstein2016kernel}, we use this correction for our tests. When the subtests are executed sequentially rather than in parallel, the procedure can terminate early if any subtest fails to reject its corresponding null, thereby saving computational time. We provide a detailed description of the test procedure of $\overline{\mathrm{x}}\mathrm{LI}$ for $d=3$ in Appendix~\ref{app: lancaster_3}.

\subsubsection{Streitberg Interaction Test}
As discussed, the Lancaster interaction is unable to detect factorisations that do not involve singletons. The measure that does vanish for all factorisations of the joint distribution is the Streitberg interaction in Eq.~\eqref{eqn: streitberg_interaction}. To test if $\mathbb{P}_{1\cdots d}$ can be factorised in any way, we have the following hypothesis:
\begin{hypothesis}(Complete Factorisation)
\label{hyp:complete_factor}
    \begin{align*}
    &H_0: \bigvee_{\pi} H_{\pi} &\forall\,|\pi| = 2 \\
    &H_1: \bigwedge_{{\pi }} \neg H_{\pi} &\forall\,|\pi| = 2
    \end{align*}
where $H_{\pi}$ denotes the event such that the joint distribution $\mathbb{P}_{1\cdots d}$ is factorised as $\mathbb{P}_{\pi}=\mathbb{P}_{b_1}\mathbb{P}_{b_2}$. 
\end{hypothesis}
Note that only partitions with exactly two blocks need to be considered in the composite tests, as all further factorisations $\mathbb{P}_{1\cdots d}=\mathbb{P}_{b_1}\mathbb{P}_{b_2}\cdots\mathbb{P}_{b_{|\pi|}}$ are subsumed in bipartition subtests. The partitions with two blocks correspond to the second level of the $d$-order partition lattice~\citep{liu2023interaction}.

The Streitberg (complete) factorisation hypothesis includes the Lancaster factorisation hypothesis plus, additionally, the $H_{\pi_S}$ where $\pi_S$ are  partitions into two blocks with no singletons.
For example, for $d=6$, the Streitberg subtests must check specifically for $\mathbb{P}_{123}\mathbb{P}_{456}$, whereas, in contrast,  $\mathbb{P}_{1}\mathbb{P}_{23456}$ is already included in the Lancaster factorisation tests (due to the singleton), and $\mathbb{P}_{12}\mathbb{P}_{34}\mathbb{P}_{56}$ does not need to be checked, since it has cardinality 3 and is subsumed in 2-block factorisations $\mathbb{P}_{1234}\mathbb{P}_{56}$, $\mathbb{P}_{1256}\mathbb{P}_{34}$ and $\mathbb{P}_{12}\mathbb{P}_{3456}$.
\begin{definition}
Under the null subhypothesis of $H_{\pi_S}$, where $\pi_S=b_1|b_2$, the permutation-free Streitberg interaction is defined as
    \begin{align*}
        \overline{\mathrm{x}}\mathrm{SI}&=\frac{\sqrt{n} \cdot\mathrm{xSI}}{s_S}\,, \quad \text{with}
        \\
        \mathrm{xSI} &= \frac{1}{n^2}\mathbf{1}_u^T\left(\overline{\bigcircle_{p\in b_1} \overline{K}^p}\circ\overline{\bigcircle_{q\in b_2} \overline{K}^{q}}\right)\mathbf{1}_\ell \,,
        \\
        s_S^2&=\frac{1}{n^3} \mathbf{1}_u^T\left\lVert\left(\overline{\bigcircle_{p\in b_1} \overline{K}^p}\circ \overline{\bigcircle_{q\in b_2} \overline{K}^{q}}\right)\mathbf{1}_\ell\right\rVert^{\circ 2} -\mathrm{xSI}^2 \, .
    \end{align*}
\end{definition}

Unlike the permutation-based Streitberg statistic, whose time complexity grows combinatorially with  $d$~\citep{liu2023interaction}, 
$\overline{\mathrm{x}}\mathrm{SI}$ grows only quadratically with $n$. 
\begin{proposition}\label{prop: xSI_time}
        The time complexity of $\overline{\mathrm{x}}\mathrm{SI}$ for all $H_{\pi_S}$ across any $d$ is $\mathcal{O}(dn^2)$.
\end{proposition}

In our numerics, we test the complete factorisation hypothesis~\eqref{hyp:complete_factor} via a two-step process we denote $\overline{\mathrm{x}}\mathrm{LI}$+$\overline{\mathrm{x}}\mathrm{SI}$: we first use $\overline{\mathrm{x}}\mathrm{LI}$ to test the $d$ subhypotheses involving $\pi_L$ (bipartitions with a singleton), followed by applying $\overline{\mathrm{x}}\mathrm{SI}$ for the subhypotheses involving the non-singleton bipartitions, $\pi_S$.  Together, these two sets cover all the $(2^{d-1}-1)$ bipartitions needed to test for the complete factorisation, i.e., $\pi_{|\pi|=2} = \pi_L \cup \pi_S$. 
Although $\overline{\mathrm{x}}\mathrm{SI}$ could, in principle, be used to perform all subtests in Hypothesis~\eqref{hyp:complete_factor} directly, we found this approach to be computationally less advantageous. 
Indeed, given that \citet{liu2023interaction} showed that the permutation-based $\mathrm{SI}$ test involves only non-singleton partitions, if we were to use $\overline{\mathrm{x}}\mathrm{SI}$, the terms in the permutation-free $\overline{\mathrm{x}}\mathrm{SI}$ exceed this scope, further contributing to its higher computational cost if it were applied for all subtests. 
As a result, using $\overline{\mathrm{x}}\mathrm{SI}$ alone does not always guarantee improved computational efficiency.
In contrast, $\overline{\mathrm{x}}\mathrm{LI}$+$\overline{\mathrm{x}}\mathrm{SI}$ ensures efficiency, by maintaining quadratic complexity with the number of samples $n$ regardless of order $d$. This advantage becomes increasingly significant as $d$ grows, making our method more suitable for practical applications involving high-order data.

\subsection{Computational Complexity}\label{sec: order}
Table~\ref{tab:order} provides a summary of the time complexity of the permutation-based and permutation-free tests, highlighting the computational advantage of the tests proposed in this paper. 
For the joint independence ($\mathrm{dHSIC}$) and Lancaster interaction ($\mathrm{LI}$) tests, the permutation-free tests offer a $p$-fold reduction in  complexity by eliminating the need for resampling.
For the complete factorisation test ($\mathrm{SI}$), our strategy of combining $\overline{\mathrm{x}}\mathrm{SI}$ and $\overline{\mathrm{x}}\mathrm{LI}$ is far more efficient than the permutation-based test in two ways: (i) $p$-fold reduction for each each subtest;  (ii) $n^2$ scaling decoupled from $d$. Empirical evaluation of these scalings is shown in Section~\ref{sec: experiments} using synthetic data with ground truths.

\begin{table}[htb!]
\centering
\caption{\textbf{Time complexity of high-order tests.} Here $p$ is the number of permutations, $n$ is the number of samples, $d$ is the number of variables and $F_d$ is the number of partitions without singletons for $d$ variables, a combinatorial number. The permutation-free tests developed in this paper are highlighted in blue.}
\vspace{0.5em}
\label{tab:order}
\begin{tabular}{lll}
\hline
\multicolumn{1}{l}{Permutation-based} & \multicolumn{1}{l}{Permutation-free} \\ \hline
$\mathrm{dHSIC}$: $\mathcal{O}(pdn^2)$ & $\textcolor{blue}{\overline{\mathrm{x}}\mathrm{dHSIC}}$: $\textcolor{blue}{\mathcal{O}(dn^2)}$ \\
\begin{tabular}[c]{@{}l@{}}
$\mathrm{LI}:\mathcal{O}(pd^2 n^2)$\\ $\mathrm{SI}$: $\mathcal{O}(p F_d^2 2^d n^{d/2})$\end{tabular} & \begin{tabular}[c]{@{}l@{}}$\textcolor{blue}{\overline{\mathrm{x}}\mathrm{LI}}$: $\textcolor{blue}{\mathcal{O}(d^2 n^2)}$\\ {$\textcolor{blue}{\overline{\mathrm{x}}\mathrm{SI}+\overline{\mathrm{x}}\mathrm{LI}}$}: \textcolor{blue}{$\mathcal{O}(2^d d n^2)$}\end{tabular} \\ \hline
\end{tabular}
\end{table}

\section{Related Work}\label{sec: related_work}
Our work is inspired by \citet{shekhar2023permutation} who introduced $\overline{\mathrm{x}}\mathrm{HSIC}$ to make $\mathrm{HSIC}$~\citep{gretton2007kernel} permutation-free by approximating HSIC using the data-splitting technique and U-statistic~\citep{kim2024dimension, shekhar2022permutation}, and normalising by the empirical standard deviation, thereby achieving a standard normal null distribution.
However, to the best of our knowledge, these methods do not extend readily to more than two variables. 
In contrast, our kernel-based formulation leverages V-statistics and cross-centring to define permutation-free tests for Lancaster and Streitberg interactions for $d>2$ variables. 

\textit{\underline{Remark:}}
For $d=2$, our formulation recovers pairwise independence (Fig.~\ref{fig: family_tree}). Yet by using cross-centring and V-statistics we obtain a permutation-free statistic $\overline{\mathrm{x}}\mathrm{HSIC}_\text{V} := \overline{\mathrm{x}}\mathrm{SI}_{d=2} = \overline{\mathrm{x}}\mathrm{LI}_{d=2}$, which is simpler, more compact, and with higher power than the original  $\overline{\mathrm{x}}\mathrm{HSIC}$ in \citet{shekhar2023permutation}. See derivation and numerics in Appendix~\ref{app: xHSIC_v}.

\section{Experiments}\label{sec: experiments}
We evaluate our proposed permutation-free tests using a range of experiments. For all cases, under the null hypothesis of each test, we confirm standard normality and controlled type-I errors of the corresponding ${\overline{\mathrm{x}}\mathrm{dHSIC}}$, $\overline{\mathrm{x}}\mathrm{LI}$ \& $\overline{\mathrm{x}}\mathrm{SI}$ statistics (see Appendix~\ref{app: experiments}, Figs.~\ref{fig: normality}--\ref{fig: type-I} for examples). 
Below we focus on the statistical power and computational efficiency of these methods compared to their permutation-based counterparts. Unless noted otherwise, 
the significance level is set to $\alpha = 0.05$, and we use Gaussian kernels with bandwidth given by the median heuristic. Additional details of each experiment are available in Appendix~\ref{app: experiments_descriptions}.

\begin{figure}[htb!]
  \centering
  \includegraphics[width=0.95\linewidth]{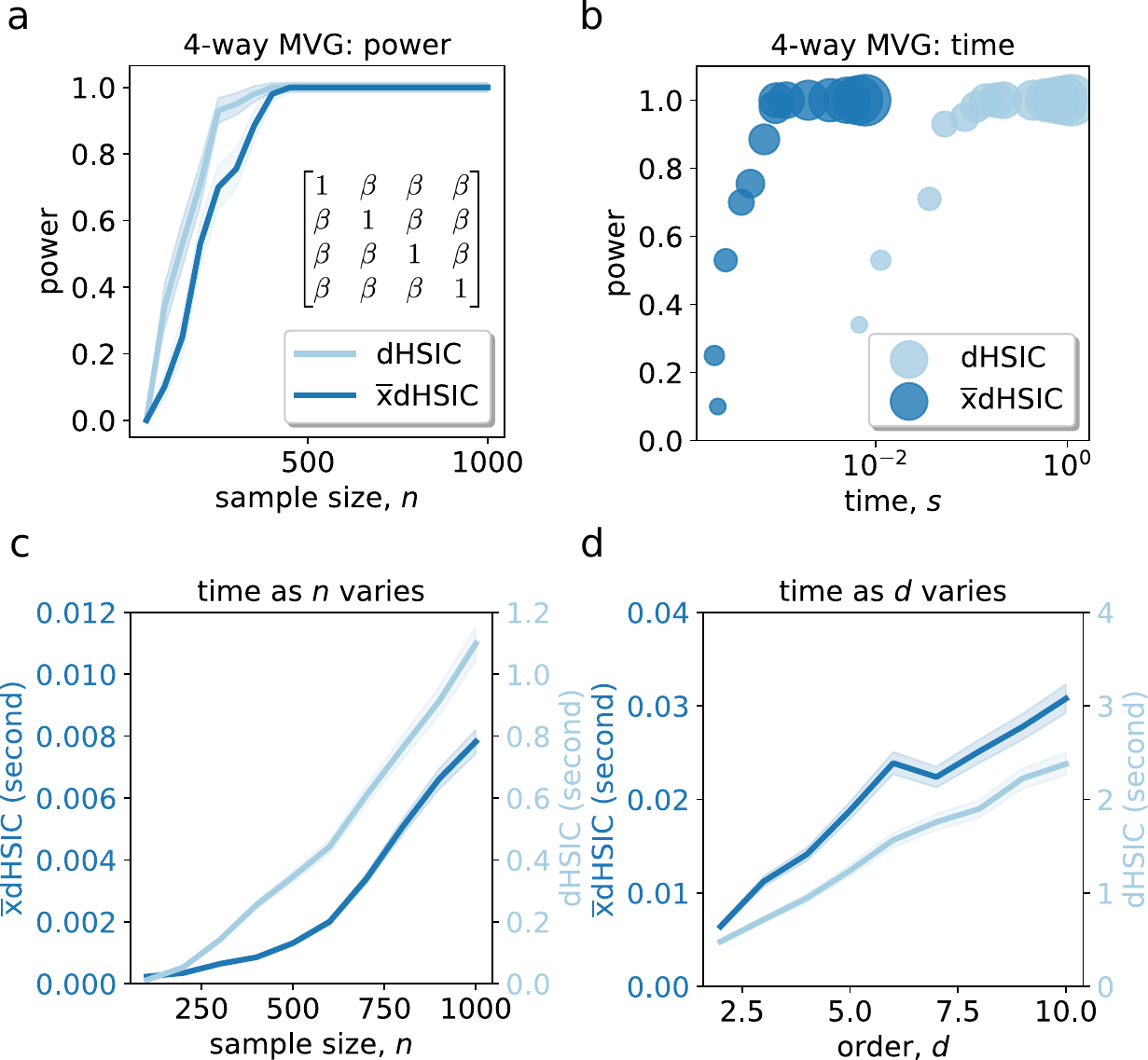}
  \caption{\textbf{Joint independence test.} (a)-(c) Sampling from a $d=4$ variable MVG (covariance matrix,  inset of (a)), we compare $\mathrm{dHSIC}$ (with $p=100$ permutations) and $\overline{\mathrm{x}}\mathrm{dHSIC}$: (a) statistical power as a function of the number of samples, $n$; (b) trade-off between power and computational time (circle diameter proportional to sample size $n$); (c) CPU time as $n$ increases. 
  (d) CPU time for MVGs of the same form as in (a) but with increasing number of variables, $d$ (with $n=500$ samples).  Note that the right scale of the CPU time in (c) and (d) is 100 times larger, thus reflecting the $p$-fold computational reduction over $\mathrm{dHSIC}$.
  }
  \label{fig: exp_xdHSIC}
\end{figure}

\subsection{Validation on Synthetic Examples}

\paragraph{Joint Independence Test}
We first demonstrate the  ${\overline{\mathrm{x}}\mathrm{dHSIC}}$ statistic for permutation-free joint independence testing.
Figure~\ref{fig: exp_xdHSIC}a-c shows the application to a $d=4$ multivariate Gaussian (MVG) distribution (covariance matrix in the inset of (a), with $\beta=0.5$). The power of the
${\overline{\mathrm{x}}\mathrm{dHSIC}}$ test has similar sensitivity compared to the permutation-based $\mathrm{dHSIC}$ test with $p=100$ permutations, but  
${\overline{\mathrm{x}}\mathrm{dHSIC}}$ reaches the maximum power roughly 100 times faster (Fig.~\ref{fig: exp_xdHSIC}b-c), as expected from the $p$-fold reduction in complexity.
The same reduction in computation time is obtained when applied to $d$-variable MVGs of varying $d$ (Fig.~\ref{fig: exp_xdHSIC}d).

\begin{figure}[htb!]
  \centering
  \includegraphics[width=0.95\linewidth]{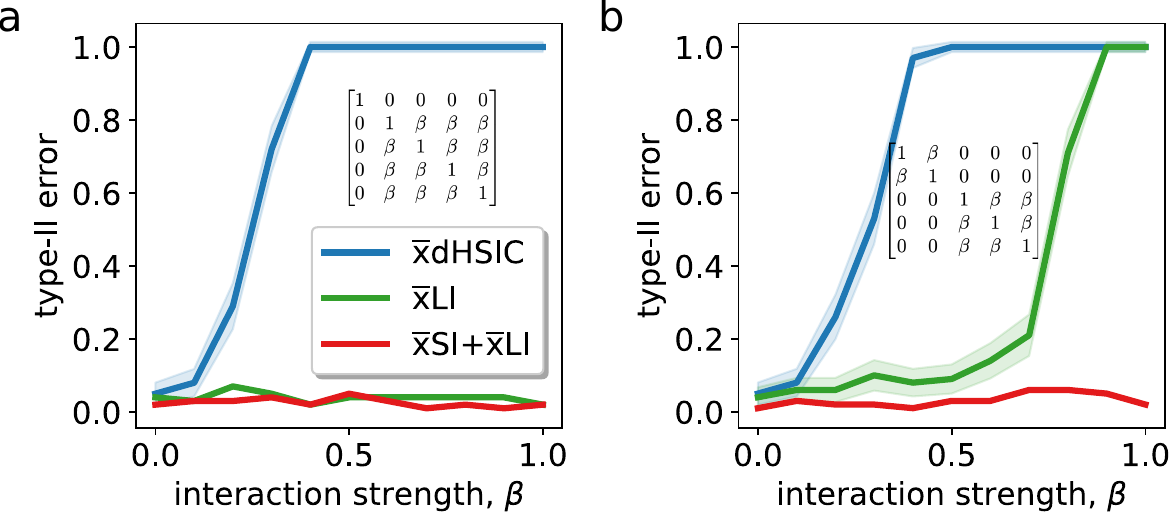}
  \caption{\textbf{Partial factorisations of MVG with $d=5$ variables.} (a)  $\overline{\mathrm{x}}\mathrm{dHSIC}$ fails to detect the partial factorisation with a singleton, $\mathbb{P}_1 \mathbb{P}_{2345}$, regardless of interaction strength. (b) Only $\overline{\mathrm{x}}\mathrm{SI}$+$\overline{\mathrm{x}}\mathrm{LI}$ identifies the factorisation with no singletons, $\mathbb{P}_{12} \mathbb{P}_{345}$. MVG covariance matrices are insets in each case.
}\label{fig: vanish}
\end{figure}

\paragraph{Factorisation Tests}
Next, we demonstrate that our permutation-free high-order tests successfully preserve the vanishing conditions for different factorisation tests. 
In Figure~\ref{fig: vanish}a, we consider data from a MVG with $d=5$ variables and covariance matrix such that $\mathbb{P}_1 \mathbb{P}_{2345}$ (see inset). As expected, $\overline{\mathrm{x}}\mathrm{dHSIC}$ fails to detect this partial (singleton) factorisation, which is correctly detected by both $\overline{\mathrm{x}}\mathrm{LI}$ and $\overline{\mathrm{x}}\mathrm{SI}$+$\overline{\mathrm{x}}\mathrm{LI}$.  
Figure~\ref{fig: vanish}b shows that, when analyse data from a MVG that is factorisable as $\mathbb{P}_{12}\mathbb{P}_{345}$ (no singletons), only $\overline{\mathrm{x}}\mathrm{SI}$+$\overline{\mathrm{x}}\mathrm{LI}$ succeeds in detecting this interaction.

\begin{figure}[htb!]
  \centering
  \includegraphics[width=0.95\linewidth]{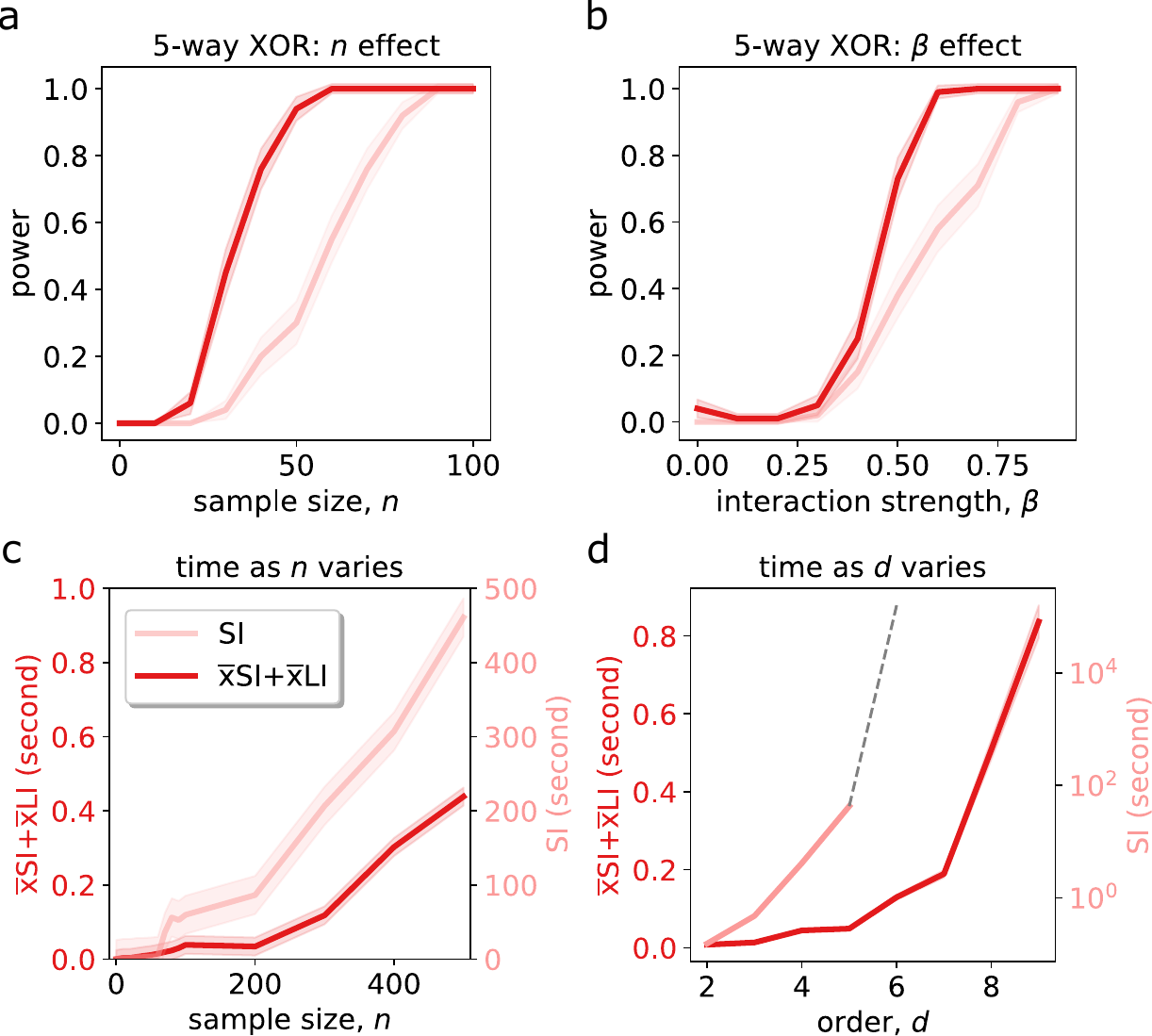}
  \caption{\textbf{5-way XOR factorisation.} Power of $\overline{\mathrm{x}}\mathrm{SI}$+$\overline{\mathrm{x}}\mathrm{LI}$ and $\mathrm{SI}$ tests for an XOR gate with $d=5$ variables as a function of: (a) sample size, $n$; (b) interaction strength, $\beta$. Computation time of both tests for: (c) increasing  sample size, $n$ (for $d=5$); (d) increasing number of variables, $d$ (for $n=100$). The $\mathrm{SI}$ test is computationally infeasible above $d=5$ (dashed grey line). Note the different left/right scales for CPU times in (c) and (d), highlighting the large reduction in computational cost achieved by $\overline{\mathrm{x}}\mathrm{SI}$+$\overline{\mathrm{x}}\mathrm{LI}$. 
}\label{fig: exp_xSI}
\end{figure}

We then simulate data from a XOR gate with $d=5$ variables, in which the interaction is exclusively of order 5 without 2-, 3-, or 4-order interactions. 
Fig.~\ref{fig: exp_xSI}a-c shows that the power of the permutation-free $\overline{\mathrm{x}}\mathrm{SI}$+$\overline{\mathrm{x}}\mathrm{LI}$ is superior to the permutation-based $\mathrm{SI}$ with a substantial reduction in computation time. 
For example, for $n=500$ samples, $\overline{\mathrm{x}}\mathrm{SI}$+$\overline{\mathrm{x}}\mathrm{LI}$ can be computed within 0.5 seconds whereas $\mathrm{SI}$ requires roughly 6 minutes. 
In Figure~\ref{fig: exp_xSI}d, we consider a series of XOR gates with increasing number of variables, $d$.  
Whereas $\overline{\mathrm{x}}\mathrm{SI}$+$\overline{\mathrm{x}}\mathrm{LI}$ can be computed in less that 1 second for $d=10$, $\mathrm{SI}$ becomes already computational infeasible for $d=6$, due to the rapid combinatorial increase in the number of non-singleton partitions ($F_d$)~\citep{sloane2018oeis}, 
which makes the computation quickly intractable as $d$ increases (Table~\ref{tab:order}).

\subsection{Applications to Causal Discovery}

\paragraph{V-Structure Detection}

A V-structure (or \emph{collider}) is a fundamental component
in constraint-based causal discovery,
where two variables $X$ and $Y$ both causally influence a third variable $Z$.
If $X\independent Y$, rejecting the Lancaster factorisation hypothesis for $d=3$ is equivalent to rejecting the conditional independence $X\independent Y\vert Z$, thereby indicating the presence of a V-structure~\cite{sejdinovic2013kernel}.

\begin{figure}[htb!]
    \centering
    \includegraphics[width=0.95\linewidth]{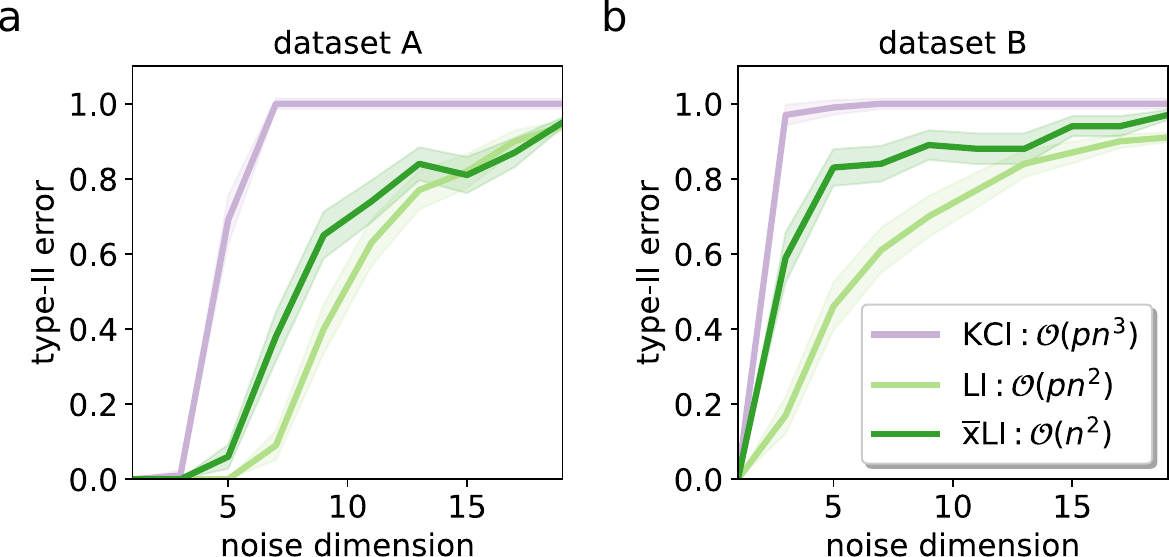}
    \caption{\textbf{V-structure detection.} Type-II errors of $\overline{\mathrm{x}}\mathrm{LI}$, $\mathrm{LI}$ and $\mathrm{KCI}$ for detecting the conditional dependence as the noise dimensions vary in dataset (a) A and (b) B from \citet{sejdinovic2013kernel}.}
    \label{fig:V-structure}
\end{figure}

Here, we use two datasets (A and B) from \citet{sejdinovic2013kernel}, in which $X\independent Y$, and we increase the noise dimension to make the causal relationship more difficult to detect.
Figures~\ref{fig:V-structure}a-b shows the comparison of our permutation-free $\overline{\mathrm{x}}\mathrm{LI}$ to the  permutation-based $\mathrm{LI}$ and kernel-based conditional independence $\mathrm{KCI}$~\citep{zhang2013Kernel_KCI} tests. 
Each experiment uses $n = 500$ samples and the 
time complexity is $\mathcal{O}(n^2)$, $\mathcal{O}(p n^2)$ and $\mathcal{O}(p n^3)$, respectively. 
$\mathrm{KCI}$ rapidly fails as the noise dimension grows, whereas Lancaster-based approaches remain more robust. 
Although $\overline{\mathrm{x}}\mathrm{LI}$ is marginally less powerful than $\mathrm{LI}$, it runs approximately 100 times faster (0.082s vs. 7.6s for $p=100$ permutations), providing a trade-off between statistical power and computational efficiency, making $\overline{\mathrm{x}}\mathrm{LI}$ an attractive choice for detecting V-structures when conventional conditional independence tests become unreliable or too slow.

Furthermore, in Appendix~\ref{app: lancaster_3} we thoroughly evaluate the sensitivity of the permutation-free method by applying it to V-structures encoded by three nonlinear forms (sinc, log, polynomial) using three different kernel functions (RQ, Laplace, RBF). We show that regardless of the non-linear form or kernel, our proposed method achieves similar accuracy to the permutation-based counterpart (Fig.~\ref{fig: vstruct_supp}).

\paragraph{DAG Causal Structural Learning}
An alternative route for causal learning is provided by score-based methods~\citep{peters2014causal,laumann2023kernel}, whereby 
each variable is modelled as a function of its direct causes (i.e., parent nodes in a directed acyclic graph (DAG)) plus an additive noise term. If the noise terms for all regressions are jointly independent, then the proposed DAG cannot be rejected.

\begin{figure}[htb!]
    \centering
    \includegraphics[width=0.95\linewidth]{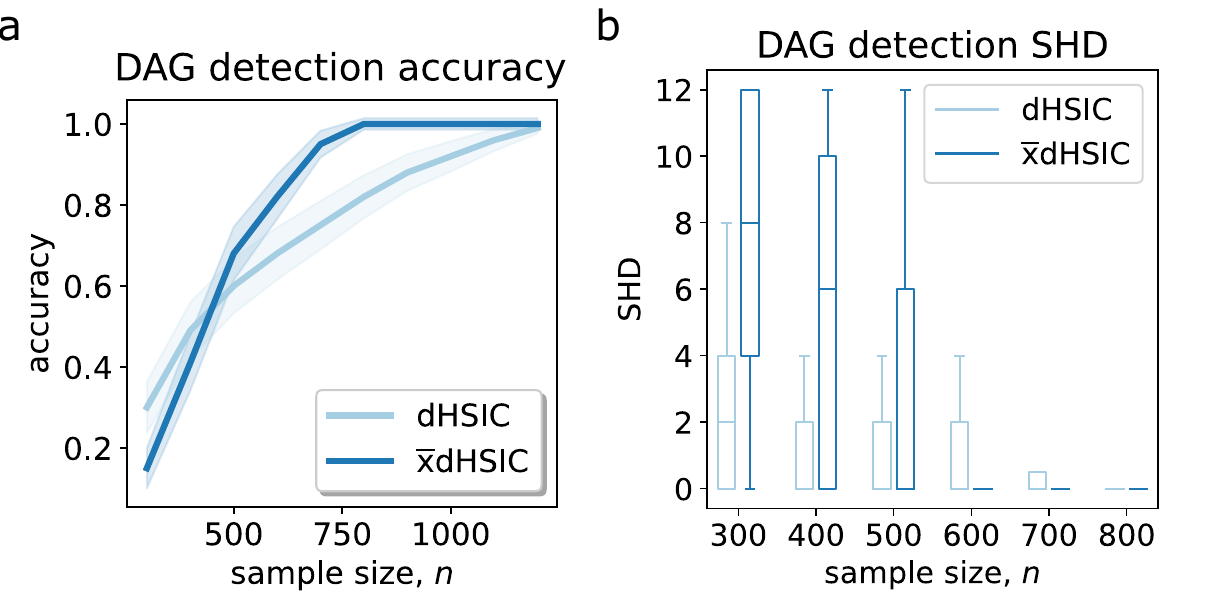}
    \caption{\textbf{DAG causal discovery.} (a) Accuracies of the correct DAG being detected among all fully connected DAGs of four nodes and (b) the distributions of the SHD of the detected DAG using $\overline{\mathrm{x}}\mathrm{dHSIC}$ and $\mathrm{dHSIC}$ out of 100 experiments.}
    \label{fig: causal}
\end{figure}

Following closely Simulation 4 from \citet{pfister2018kernel}, we simulate $n$ samples from one DAG with $d=4$ variables (see Appendix~\ref{app: experiments}). We then compare the accuracy of our permutation-free $\overline{\mathrm{x}}\mathrm{dHSIC}$ against the permutation-based $\mathrm{dHSIC}$ (using $p=200$) to recover the ground truth DAG.
Figure~\ref{fig: causal}a shows that $\overline{\mathrm{x}}\mathrm{dHSIC}$ identifies the correct DAG more frequently than  $\mathrm{dHSIC}$ for $n>500$, and it 
reaches perfect accuracy using roughly two-thirds of the samples needed by $\mathrm{dHSIC}$, all of it with a 200-fold computational speed-up. 
Similarly, the structural Hamming distance (SHD) of $\overline{\mathrm{x}}\mathrm{dHSIC}$ reaches zero faster than $\mathrm{dHSIC}$ (Fig.~\ref{fig: causal}b).

In real-world causal discovery, one often needs to search over all possible DAGs (see Appendix~\ref{app: experiments}). For 3, 4, 5, and 6 nodes, there are 25, 543, 29281, and 3781503 DAGs, respectively. Since the ratio of DAG counts between consecutive node sets $\{3, 4, 5, 6\}$ is less than 200, the computation time required for $\mathrm{dHSIC}$ (with $p=200$ permutations) to test joint independence in DAGs of 3 (or 4,5) nodes exceeds that of $\overline{\mathrm{x}}\mathrm{dHSIC}$ applied to DAGs with an additional node, i.e., 4 (or 5, 6). This efficiency gain can facilitate experiments to uncover causal relationships more effectively.

\subsection{Application to Feature Selection}
Feature selection is a crucial step in many machine learning pipelines. The task is to identify the most informative features to enhance model performance and interpretability. 
Most techniques, however, focus solely on \emph{pairwise} relationships between a feature and the target variable~\citep{poczos2012copula, song2012feature}, overlooking high-order interactions~\citep{fumagalli2024unifying,muschalik2024shapiq}.

To illustrate the application to feature selection, we construct a dataset ($n=500$) with nine features $\{X^1, \ldots, X^9\}$, of which $\{X^1, \ldots, X^7\}$ are $i.i.d.$ uniform variables interacting through an XOR gate to generate a target variable $Y$, and $\{X^8, X^9\}$ are MVG's with covariance $[[1, 0.9], [0.9, 1]]$. The ground truth interaction between the variables is thus $\{Y, X^1,\ldots, X^7\}\independent\{X^8, X^9\}$; hence no single variable or pair of variables is predictive of $Y$.
In Table~\ref{tab: feature_selection}, we compare the performance of feature selection methods based on high-order tests with popular methods based on univariate and pairwise measures, including the sequential method, recurrent feature selection (RFE)~\cite{sklearn} and SHAP~\cite{SHAP}.
The permutation-free test $\overline{\mathrm{x}}\mathrm{SI}$+$\overline{\mathrm{x}}\mathrm{LI}$ is the only method able to detect the ground truth feature set with lowest mean squared error (MSE).
All other methods, including pairwise independence, joint independence and Lancaster tests, suffer from errors in selecting the correct features due to their inability to identify the complex high-order interactions between the input features and the target variable in this example.

\begin{table}[t]
\centering
\caption{\textbf{Feature selection.} Only $\overline{\mathrm{x}}\mathrm{SI}$+$\overline{\mathrm{x}}\mathrm{LI}$ identifies the ground truth variables ($\CIRCLE$) that generate $Y$. Other methods either include confounding variables ($\red{\CIRCLE}$) or miss generating variables ($\red{\Circle}$).}
\vspace{0.5em}
\label{tab: feature_selection}
\resizebox{0.48\textwidth}{!}{%
\begin{tabular}{lcccccccllc}
\hline
Method & $X^1$ & $X^2$ & $X^3$ & $X^4$ & $X^5$ & $X^6$ & $X^7$ & \multicolumn{1}{c}{$X^8$} & \multicolumn{1}{c}{$X^9$} & MSE \\ \hline
Variance & $\CIRCLE$ & $\CIRCLE$ & $\CIRCLE$ & $\CIRCLE$ & $\CIRCLE$ & $\CIRCLE$ & $\CIRCLE$ & $\red{\CIRCLE}$ & $\red{\CIRCLE}$ & 5.01 \\
Univariate & $\CIRCLE$ & $\CIRCLE$ & $\CIRCLE$ & $\CIRCLE$ & $\CIRCLE$ & $\CIRCLE$ & $\CIRCLE$ & $\red{\CIRCLE}$ & $\red{\CIRCLE}$ & 5.01 \\
Sequential & $\CIRCLE$ & $\CIRCLE$ & $\CIRCLE$ & $\CIRCLE$ & $\red{\Circle}$ & $\red{\Circle}$ & $\red{\Circle}$ & $\red{\CIRCLE}$ &  & 4.69 \\
RFE & $\CIRCLE$ & $\CIRCLE$ & $\CIRCLE$ & $\red{\Circle}$ & $\red{\Circle}$ & $\red{\Circle}$ & $\red{\Circle}$ & $\red{\CIRCLE}$ &  & 4.47 \\
SHAP & $\CIRCLE$ & $\CIRCLE$ & $\CIRCLE$ & $\CIRCLE$ & $\CIRCLE$ & $\red{\Circle}$ & $\red{\Circle}$ & $\red{\CIRCLE}$ & $\red{\CIRCLE}$ & 4.67 \\
$\overline{\mathrm{x}}\mathrm{HSIC}$ & $\CIRCLE$ & $\CIRCLE$ & $\CIRCLE$ & $\CIRCLE$ & $\CIRCLE$ & $\red{\Circle}$ & $\red{\Circle}$ & $\red{\CIRCLE}$ & $\red{\CIRCLE}$ & 4.67 \\
$\overline{\mathrm{x}}\mathrm{dHSIC}$ & $\CIRCLE$ & $\CIRCLE$ & $\CIRCLE$ & $\CIRCLE$ & $\CIRCLE$ & $\CIRCLE$ & $\CIRCLE$ & $\red{\CIRCLE}$ & $\red{\CIRCLE}$ & 5.01 \\
$\overline{\mathrm{x}}\mathrm{LI}$ & $\CIRCLE$ & $\CIRCLE$ & $\CIRCLE$ & $\CIRCLE$ & $\CIRCLE$ & $\CIRCLE$ & $\CIRCLE$ & $\red{\CIRCLE}$ & $\red{\CIRCLE}$ & 5.01 \\
\vspace*{-.12in}
\\
$\overline{\mathrm{x}}\mathrm{SI}$+$\overline{\mathrm{x}}\mathrm{LI}$ & $\CIRCLE$ & $\CIRCLE$ & $\CIRCLE$ & $\CIRCLE$ & $\CIRCLE$ & $\CIRCLE$ & $\CIRCLE$ &  &  & \textbf{4.35} \\
\hdashline
\textit{Ground truth} & $\CIRCLE$ & $\CIRCLE$ & $\CIRCLE$ & $\CIRCLE$ & $\CIRCLE$ & $\CIRCLE$ & $\CIRCLE$ &  &  & \textit{4.35} \\ \hline
\end{tabular}%
}
\end{table}

\subsection{Application to Dataset of Stock Daily Returns}
\begin{figure*}[t]
    \centering
    \includegraphics[width=0.9\linewidth]{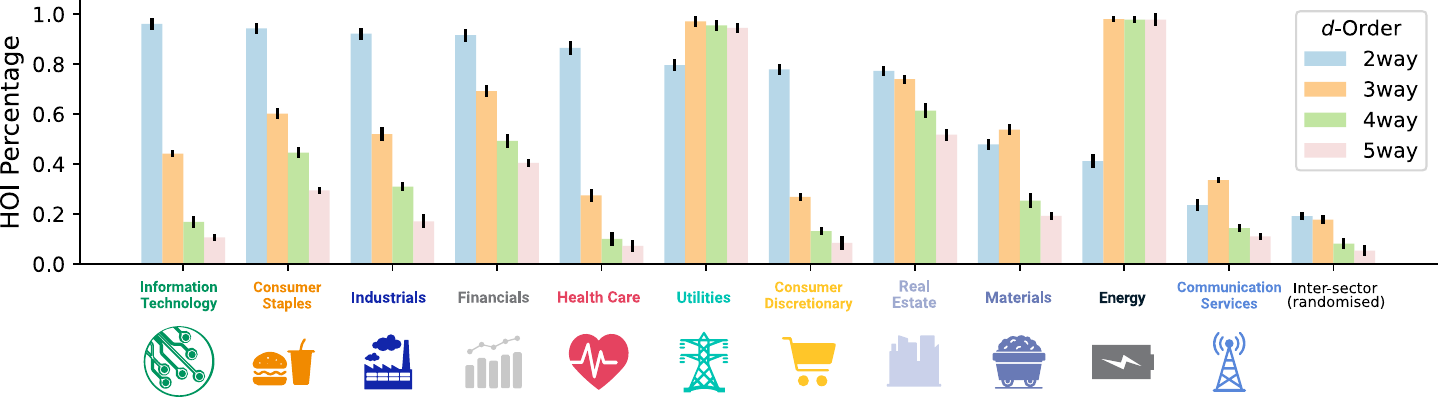}
    \caption{\textbf{Stock interactions in S\&P500.} Utilities and Energy show high 3-, 4– and 5-way interactions, likely due to regulation-driven redundancy. In contrast, Information Technology, Health Care, and Consumer Discretionary exhibit lower high-order interactions, suggesting stronger cross-sector connectivity.}
    \label{fig: sp500}
\end{figure*}
To demonstrate the scalability and applicability of our permutation-free tests, we consider a large real-world financial dataset: daily returns of stocks in the S\&P 500 from 2020 to 2024. This dataset comprises 504 variables (stocks) and 1004 samples (returns), spanning 11 sectors (Fig.~\ref{fig: sp500}).
As is standard in the finance literature, we assume the daily returns to be $i.i.d.$~\cite{ali1982identical}.

Figure~\ref{fig: sp500} shows the percentage of  2-, 3-, 4-, and 5-way interactions for all sectors, where we compute interactions for 500 sets of stocks within the same sector and, as a comparison, for 5000 sets of stocks randomly drawn from different sectors. 
As expected, 2-, 3-, 4-, and 5-way interactions are more common within a sector than across sectors, in concordance with the GICS industrial taxonomy that underpins the definition of the sectors. Interestingly, both Utilities and Energy exhibit very high 3-, 4-, and 5-way within-sector interactions. This likely stems from the highly regulated nature of these sectors, resulting in stocks with similar returns and, consequently, high (within-sector) redundancy. Conversely, Information Technology, Consumer Discretionary and Health Care display lower high-order interactions. This suggests that companies within these sectors may be more closely connected to firms in other sectors as they are likely influenced by external factors, indicating (cross-sector) synergistic relationships. A summary heatmap of the high-order interaction profiles of all sectors is shown in  Figure~\ref{fig: sp500_heatmap} in Appendix~\ref{app: experiments_descriptions}. 
These high-order interactions could be of help when considering portfolio diversification.

\section{Discussion}\label{sec: discussion}

The use of fine-tuned permutation schemes has been shown to be an essential ingredient for interaction tests to produce reliable results, but this process can be both time-consuming and challenging~\cite{rindt2021consistency}. The permutation-free methods introduced here, $\overline{\mathrm{x}}\mathrm{dHSIC}$, $\overline{\mathrm{x}}\mathrm{LI}$ and $\overline{\mathrm{x}}\mathrm{SI}$, offer a more efficient approach to handling complex data structures circumventing the need for permutations and the tuning of associated hyperparameters, thus simplifying the workflow and ensuring consistent performance.

Regarding when to use permutation-free vs.\ permutation-based tests, our experiments show that for small sample size, permutation-based methods perform comparably, and occasionally slightly better. 
This is likely due to the need for data splitting in permutation-free methods, which limits their ability to capture the full data structure in low-sample regimes as test statistics rely on inner products between embeddings estimated from the two separate halves. Hence, as a rule of thumb, for low number of samples $n$ and low order $d$ the trade-off between computation time and statistical power would not be unfavourable for permutation-based methods, e.g., for $n\leq 50$ and $d\leq 5$ permutation-based methods perform well with just $p=100$ permutations. Outside this regime, however, we strongly recommend the permutation-free methods introduced here.

Nevertheless, several challenges and limitations remain. The data-splitting technique that underpins our permutation-free strategy relies on the assumption of $i.i.d.$ samples, which can be restrictive when dependencies exist across samples (e.g., time series, network data, spatial data). Although permutation-based techniques have been developed to address stationary~\citep{chwialkowski2014kernel, rubenstein2016kernel} and non-stationary time-series data~\citep{liu2023kernel}, the development of permutation-free methods for such data remains an open challenge. Additionally, while we focus here on joint independence and factorisation hypotheses, other forms of higher-order structure, such as conditional independence or more intricate composite hypotheses, might benefit from similar permutation-free schemes. Finally, although the permutation-free approach substantially improves the efficiency of high-order tests, there remains potential to further reduce computational overhead by incorporating scalable kernel approximation techniques such as random Fourier features and the Nystr\"om method~\cite{zhang2018large}. Investigating these directions could further broaden the applicability of kernel-based high-order interaction tests in large-scale machine learning and statistical settings.

\paragraph{Code Availability}
Code to implement the permutation-free tests in this paper available at
\url{https://github.com/barahona-research-group/PermFree-HOI.git}.

\section*{Acknowledgements}
MB acknowledges support by EPSRC through grants EP/W024020/1, funding the project ``Statistical physics of cognition'', and EP/N014529/1, funding the EPSRC Centre for Mathematics of Precision Healthcare, and by the Nuffield Foundation, under the project ``The Future of Work and Well-being: The Pissarides Review''.
RP acknowledges funding from the Deutsche Forschungsgemeinschaft (DFG) Project-ID 424778381-TRR 295.

The authors thank Felix Laumman, Jianxiong Sun, Shubhanshu Shekhar and Zhuoyu Li for valuable discussions.

\section*{Impact Statement}

The societal impact of our work stems from its ability to address the critical need for efficient and scalable methods to detect high-order interactions in multivariate data. By eliminating the reliance on computationally intensive permutation schemes, our permutation-free tests enable researchers to uncover complex relationships beyond pairwise interactions, which are crucial in domains like causal discovery and feature selection. These advancements facilitate more effective data analysis in fields such as healthcare, economics, and environmental science, where understanding intricate interactions can lead to breakthroughs in diagnosis, decision-making, and policy formulation.
\bibliographystyle{icml2025}
\bibliography{ref}

\newpage
\appendix
\onecolumn

\section*{Appendices}

\section{Proofs for Section~\ref{sec: permfree_highorder}}\label{app: proof}

\subsection{Proof of Proposition~\ref{prop: xdHSIC_time}}
First observe that each term in the numerator can be computed in $\mathcal{O}(dn^2)$ since the expression only contains Hadamard product and vector matrix multiplication~\citet{pfister2018kernel}.
Similarly for the denominator, the variance estimation introduces no significant computational overhead beyond simple vector norm compared to $\mathrm{xdHSIC}$.

\subsection{Proof of Lemma~\ref{lem: cross-centring}}
The conventional centring of a matrix $K$ is denoted by $\tilde{K} = CKC$, where $C = I - \frac{1}{n}\textbf{1}\textbf{1}^T$ and $\textbf{1}$ is an $n\times 1$ vector of ones. It has been shown that the respective centred kernel function $\tilde{k}$ is an equivalent kernel to $k$. Indeed, as shown in \citet{sejdinovic2013equivalence}, a kernel $\tilde{k}_f$ is equivalent to $k$ if it can expressed as:
\begin{align*} 
\tilde{k}_f\left(z, z^{\prime}\right) =\left\langle k(\cdot, z)-f, k\left(\cdot, z^{\prime}\right)-f\right\rangle_{\mathrm{HS}}
\end{align*}
for some function $f$. Now for conventional centring $\tilde{k}$, $f=\mu_{\mathbb{P}_X}$ which can be approximated by $\widehat{\mu}_{\mathbb{P}_X}=\frac{1}{n} \sum_{i=1}^n \phi\left(x_i\right)$ and thus is an equivalent kernel.

Here, the cross-centred $\overline{k}$ is also equivalent as $\frac{1}{n} \sum_{i=1}^n \phi\left(x_i\right)$ and $\frac{1}{n} \sum_{i=n+1}^{2n} \phi\left(x_i\right)$ can be seen as approximation of $\mu_{\mathbb{P}_X}$.

\subsection{Proof of Proposition~\ref{prop: xLI_time}}
Here, consider the Hadamard product of all the kernels in $D$ except $K^k$ as an individual kernel, then the $\overline{\mathrm{x}}\mathrm{LI}$ can be seen as a special $\overline{\mathrm{x}}\mathrm{HSIC}$ and therefore they share the same time complexity. 

\subsection{Proof of Proposition~\ref{prop: xSI_time}}
Here, consider the two Hadamard products as two kernels themselves, then the $\overline{\mathrm{x}}\mathrm{SI}$ can be seen as a special $\overline{\mathrm{x}}\mathrm{HSIC}$ and therefore they share the same time complexity. 


\section{Extended analysis of the $d=3$ Lancaster interaction test}\label{app: lancaster_3}
\subsection{Lancaster interaction hypothesis and test}
$\Delta^3_L \mathbb{P}$ in Equation~\eqref{eqn: lancaster_3} is the extension as $d$ increases from 2 to 3 in the direction of the factorisation route as illustrated in Figure~\ref{fig: family_tree}. Now instead of testing for joint independence through the full factorisation $\mathbb{P}_{123}=\mathbb{P}_1\mathbb{P}_2\mathbb{P}_3$, we would like to understand if the joint distribution $\mathbb{P}_{123}$ is partially factorisable as summarised in the composite hypothesis below:
\begin{hypothesis} (Order-3 Factorisation)
    \begin{align*}
       H_0: 
       \left(\mathbb{P}_{123}=\mathbb{P}_{12}\mathbb{P}_{3} \right) 
       \vee
       \left(\mathbb{P}_{123}=\mathbb{P}_{13}\mathbb{P}_{2} \right) 
       \vee
       \left(\mathbb{P}_{123}=\mathbb{P}_{23}\mathbb{P}_{1} 
       \right) 
       \\
    H_1: 
    \left(\mathbb{P}_{123}\neq\mathbb{P}_{12}\mathbb{P}_{3} \right)
    \wedge
    \left(\mathbb{P}_{123}\neq\mathbb{P}_{13}\mathbb{P}_{2} \right)
    \wedge
    \left(\mathbb{P}_{123}\neq\mathbb{P}_{23}\mathbb{P}_{1} \right)
\end{align*}
\end{hypothesis}
If any of the subhypotheses in the null hypothesis $H_0$ is true, then $\Delta^3_L \mathbb{P} = 0$. For example, if $\mathbb{P}_{123} = \mathbb{P}_{12}\mathbb{P}_{3}$, this implies $\mathbb{P}_{13} = \mathbb{P}_{1}\mathbb{P}_{3}$ and $\mathbb{P}_{23} = \mathbb{P}_{2}\mathbb{P}_{3}$, and, as a result, we have:
\begin{align*}
\mathbb{P}_{123} = \mathbb{P}_{12}\mathbb{P}_{3} \implies 
    \Delta^3_L \mathbb{P} &= \mathbb{P}_{123} - \mathbb{P}_{12}\mathbb{P}_3 - \mathbb{P}_{13}\mathbb{P}_2 - \mathbb{P}_{23}\mathbb{P}_{1} +2\mathbb{P}_{1}\mathbb{P}_2\mathbb{P}_3  = 0
\end{align*}
And, similarly, $\mathbb{P}_{123} = \mathbb{P}_{13}\mathbb{P}_{2} \implies \Delta^3_L \mathbb{P}=0$ and $\mathbb{P}_{123} = \mathbb{P}_{23}\mathbb{P}_{1} \implies \Delta^3_L \mathbb{P}=0$.  Hence $H_0 \implies \Delta^3_L \mathbb{P}=0$.
Note that even though $H_0$ only considers the three partial factorisations, the factorisation $\mathbb{P}_1\mathbb{P}_2\mathbb{P}_3$ is actually subsumed by them, i.e. $\mathbb{P}_{123} = \mathbb{P}_1\mathbb{P}_2\mathbb{P}_3$ implies that $\mathbb{P}_{123} = \mathbb{P}_{12}\mathbb{P}_{3}$ and $\mathbb{P}_{123} = \mathbb{P}_{13}\mathbb{P}_{2}$ and $\mathbb{P}_{123} = \mathbb{P}_{23}\mathbb{P}_{1}$.  In other words, $\mathbb{P}_{123} = \mathbb{P}_1\mathbb{P}_2\mathbb{P}_3 \implies \Delta^3_L \mathbb{P}=0$ also.

By the contrapositive, this means:
$\Delta^3_L \mathbb{P} \neq 0 \implies \neg H_0 \equiv H_1$.
Hence, when a $d=3$ interaction is detected
via the alternative hypothesis $H_1$, all three subhypotheses in $H_0$, as well as the complete factorisation $\mathbb{P}_{123} = \mathbb{P}_1\mathbb{P}_2\mathbb{P}_3$, need to be rejected.

\subsection{Construction of the permutation-free normalised test statistic}
By computing the RKHS embedding and applying the permutation-free scheme, the unnormalised test statistics of three-variable Lancaster interaction can be immediately written out in terms of the cross-centred kernels:
\begin{align*}
    \mathrm{xLI} = \frac{1}{n^2}\sum_{i=1}^{n}{\sum_{j=n+1}^{2n}} \left[\overline{K}^1 \circ \overline{K}^2 \circ \overline{K}^3\right]_{ij}
\end{align*}
Note that a key distinction between the permutation-based methods and the permutation-free approaches discussed here lies in how multiple tests are conducted. In permutation-based methods, the test statistic remains constant, while the permutations adapt to changes in the subhypotheses. In contrast, with the permutation-free approach, where the null distributions are all standardised as $\mathcal{N}(0, 1)$, the test statistics must instead vary.

Without loss of generality, under the subhypothesis $\mathbb{P}_{123}=\mathbb{P}_{12}\mathbb{P}_{3}$, $\mathrm{xLI}$ can be simplified into:
\begin{align*}
    \mathrm{xLI}_3 = \frac{1}{n^2}\sum_{i=1}^{n}{\sum_{j=n+1}^{2n}} \left[\overline{\overline{K}^1 \circ \overline{K}^2} \circ \overline{K}^3\right]_{ij}
\end{align*}

The normalised test statistic is
\begin{align*}
    \overline{\mathrm{x}}\mathrm{LI}_3&=\frac{\sqrt{n} \cdot \mathrm{xLI}_3}{s_L}, \quad
    \text{with}\\
    s_L^2&=\frac{1}{n} \sum_{i=1}^n\left(\frac{1}{n} \sum_{j=n+1}^{2n} \left[\overline{\overline{K}^1 \circ \overline{K}^2}\circ \overline{K}^3\right]_{ij}-\mathrm{xLI}_3\right)^2\\
    &=\frac{1}{n} \sum_{i=1}^n\left(\left(\frac{1}{n} \sum_{j=n+1}^{2n} \left[\overline{\overline{K}^1 \circ \overline{K}^2}\circ \overline{K}^3\right]_{ij}\right)^2 + \mathrm{xLI}_3^2 - 2\mathrm{xLI}_3\left(\frac{1}{n} \sum_{j=n+1}^{2n} \left[\overline{\overline{K}^1 \circ \overline{K}^2}\circ \overline{K}^3\right]_{ij}\right)\right)\\
    &=\frac{1}{n} \sum_{i=1}^n\left(\frac{1}{n} \sum_{j=n+1}^{2n} \left[\overline{\overline{K}^1 \circ \overline{K}^2}\circ \overline{K}^3\right]_{ij}\right)^2 +\mathrm{xLI}_3^2-2\mathrm{xLI}_3^2\\
    & = \frac{1}{n} \sum_{i=1}^n\left(\frac{1}{n} \sum_{j=n+1}^{2n} \left[\overline{\overline{K}^1 \circ \overline{K}^2}\circ \overline{K}^3\right]_{ij}\right)^2 -\mathrm{xLI}_3^2
\end{align*}
Similarly, for the other two subhypotheses, we just move the double cross-centring to the pairwise variables. 
The whole testing procedure is summarised in Algorithm~\ref{alg: 1}. 
\begin{algorithm}[ht]
   \caption{Lancaster Interaction Test at $d=3$}
   \label{alg: 1}
\begin{algorithmic}
   \STATE {\bfseries Input:} Centred kernel matrices $\overline{K}^1$, $\overline{K}^2$, $\overline{K}^3$ computed from observational data 
   \FOR{each subhypothesis (total three subhypotheses)}
       \STATE Compute the test statistic $\mathrm{\overline{x}LI}$
       \IF{$\mathrm{\overline{x}LI} > (1-\alpha)$-quantile of $\mathcal{N}(0, 1)$}
           \STATE Reject $H_i$ and proceed to the next subhypothesis
       \ELSE
           \STATE Terminate and do not reject composite hypothesis
       \ENDIF
   \ENDFOR
   \IF{all three subhypotheses are rejected}
       \STATE Reject the composite hypothesis
   \ELSE
       \STATE Do not reject the composite hypothesis
   \ENDIF
\end{algorithmic}
\end{algorithm}

The composite hypothesis $H_0$ is rejected only if all subhypotheses are individually rejected. As this has been proven to be less conservative than the Bonferroni correction~\cite{rubenstein2016kernel}, we use this correction for the factorisation tests for any $d$. When the subtests are executed sequentially rather than in parallel, the procedure can terminate early if any subtest fails to reject its corresponding subhypothesis, thereby saving computational time.

\subsection{Application to additional examples: nonlinear form and different kernels}
In addition to the examples from \citet{sejdinovic2013kernel}, we also apply our method to three more examples with diverse nonlinear forms. In all cases, our proposed permutation-free method achieves similar accuracy compared with the permutation-based counterpart. Moreover the results are robust with different kernel functions, Rational-Quadratic (RQ) kernel, Laplace Kernel and Radial Basis Function (RBF) kernel.

\begin{figure*}[ht]
  \centering
  \includegraphics[width=0.75\linewidth]{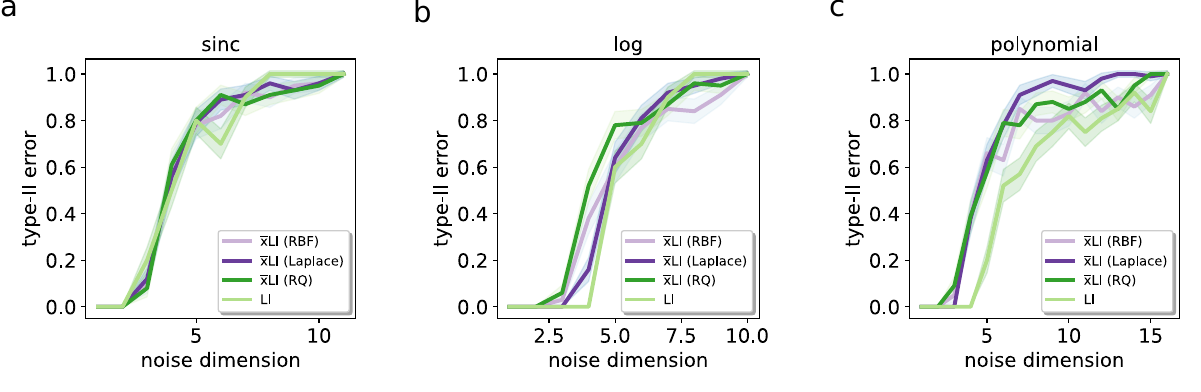}
  \caption{\textbf{Structural causal models with diverse nonlinear forms and kernel functions.} We constructed three V-structures using (a) sinc (b) log, and (c) polynomial nonlinear forms (see inset). }\label{fig: vstruct_supp}
\end{figure*}
\section{Definition of the $\overline{\mathrm{x}}\mathrm{HSIC}$ pairwise test using V-statistics}\label{app: xHSIC_v}
We formulate the corresponding V-statistic version of the permutation-free pairwise independence test statistic, 
$\overline{\mathrm{x}}\mathrm{HSIC}_\text{V}$ using the cross-centring technique:
\begin{align*}
    \overline{\mathrm{x}}\mathrm{HSIC}_\text{V} &= \frac{\sqrt{n}\cdot\mathrm{xHSIC}_\text{V}}{s_\text{V}}
    \\
    \mathrm{xHSIC}_\text{V} &= \frac{1}{n^2}\sum_{i=1}^{n}{\sum_{j=n+1}^{2n}} \left[\overline{K}^1 \circ \overline{K}^2\right]_{ij}
    \\
    s_\text{V}^2&=\frac{1}{n} \sum_{i=1}^n\left(\frac{1}{n} \sum_{j=n+1}^{2n} \left[\overline{K}^1\circ \overline{K}^2\right]_{ij}-\mathrm{xHSIC}_\text{V}\right)^2
\end{align*}
This is a simpler formulation compared with the original $\overline{\mathrm{x}}\mathrm{HSIC}$ in~\citet{shekhar2023permutation}. 

\begin{figure*}[ht]
  \centering
  \includegraphics[width=0.65\linewidth]{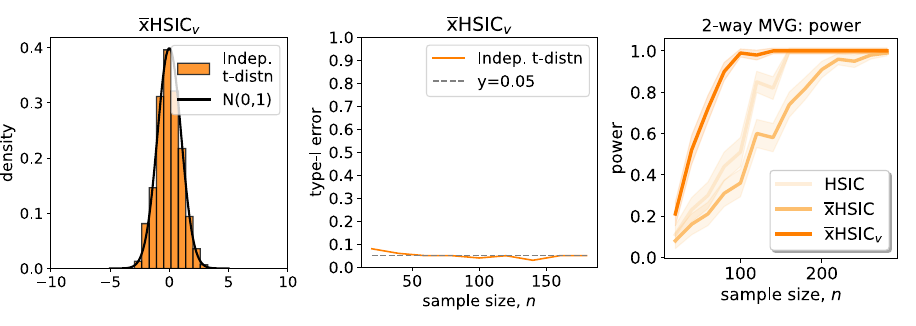}
  \caption{\textbf{Experiments of pairwise independence tests:} $\overline{\mathrm{x}}\mathrm{HSIC}_\text{V}$ follows a standard normal distribution with two independent $t$-distributions; controls the type-I error rate with the same data from $t$-distributions, and outperforms both $\mathrm{HSIC}$~\cite{gretton2007kernel} and $\overline{\mathrm{x}}\mathrm{HSIC}$~\cite{shekhar2023permutation} with dependent MVG from the example in Equation (13) from \citet{shekhar2023permutation} with parameters $b = 2$, $ndim = 10$, and $\epsilon = 0.5$.}\label{fig: exp_d2}
\end{figure*}

Figure~\ref{fig: exp_d2} shows that not only is $\overline{\mathrm{x}}\mathrm{HSIC}_\text{V}$ theoretically grounded, like $\mathrm{xHSIC}$, but it also has higher power compared with both $\overline{\mathrm{x}}\mathrm{HSIC}$ and $\mathrm{HSIC}$.

\section{Further numerical checks of the test statistics }\label{app: experiments}

\begin{figure}[H]
    \centering
    \includegraphics[width=0.65\linewidth]{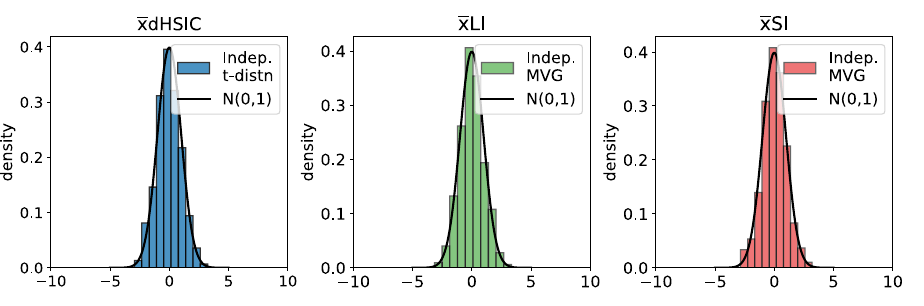}
    \caption{\textbf{Null distributions of permutation-free test statistics.} All three test statistics introduced in this paper follow a standard normal distribution $\mathcal{N}(0, 1)$ under their respective nulls.}
    \label{fig: normality}
\end{figure}

\begin{figure}[H]
    \centering
    \includegraphics[width=0.65\linewidth]{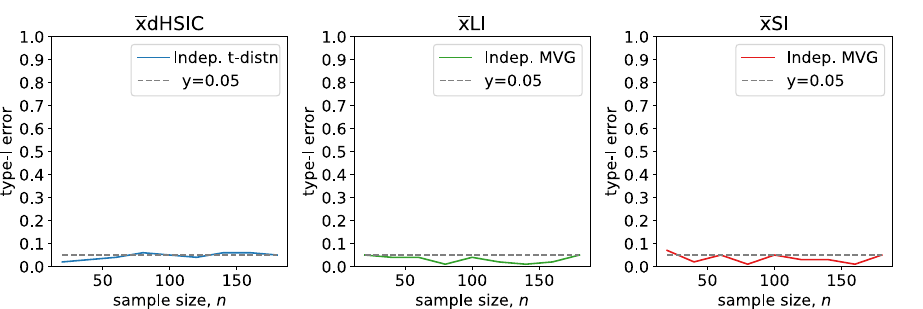}
    \caption{\textbf{Controlled type-I error of permutation-free test statistics.} All three test statistics introduced in this paper have a valid level of type-I error for $\alpha=0.05$, indicated by the dashed line.}
    \label{fig: type-I}
\end{figure}

\section{Further details of datasets used in Section~\ref{sec: experiments}}
\label{app: experiments_descriptions}
The code to implement the permutation-free tests introduced in this paper is available at \url{https://github.com/barahona-research-group/PermFree-HOI.git}.
\subsection{XOR example from \citet{liu2023interaction}}
We generate $n$ samples of $V, W, X, Y, Z \sim \mathcal{U}(0,4)$, $Z_{: i}=\left(V_{:i}+W_{: i}+X_{: i}+Y_{: i}\right) \bmod 4$ and $\left.Z_{i+1: n} \sim \mathcal{U}(0,4)\right)$ (where the samples $[i+1: n]$ act as noise). We then gradually increase the interaction proportion, $0 \leq i / n \leq 1$.
\subsection{V-structure dataset A and B from \citet{sejdinovic2013kernel}}
\textbf{Dataset A:} For $\{X, Y, Z\}\in\mathbb{R}^a \times \mathbb{R}^a \times \mathbb{R}^a$, $X, Y \sim \mathcal{N}\left(0, I_p\right)$, $W \sim \operatorname{Exp}\left(\frac{1}{\sqrt{2}}\right)$, $Z_1=\operatorname{sign}\left(X_1 Y_1\right) W$, and $Z_{2: p} \sim \mathcal{N}\left(0, I_{p-1}\right)$.

\textbf{Dataset B:} For $\{X, Y, Z\}\in\mathbb{R}^a \times \mathbb{R}^a \times \mathbb{R}^a$, $X, Y {\sim} \mathcal{N}\left(0, I_p\right), Z_{2: p} \sim \mathcal{N}\left(0, I_{p-1}\right)$, and
\begin{equation*}
    Z_1= \begin{cases}X_1^2+\epsilon, & \text { w.p. } 1 / 3, \\ Y_1^2+\epsilon, & \text { w.p. } 1 / 3 \\ X_1 Y_1+\epsilon, & \text { w.p. } 1 / 3\end{cases}
\end{equation*}
where $\epsilon\sim \mathcal{N}(0, 0.01)$ and W.P. stands for with probability. 

In both datasets, the 3-way interaction becomes increasingly difficult to detect as the noise dimension $p$ increases.
\subsection{Causal discovery example from \citet{pfister2018kernel}}
For an additive noise model over random variables $X^1, \ldots, X^d$,
\begin{equation}
    X^j:=\sum_{k \in \mathbf{Parents}} f^{j, k}\left(X^k\right)+N^j, \quad j \in\{1, \ldots, d\}
\end{equation}
with corresponding DAG $G$. The noise variables $N^j$ are normally distributed and are jointly independent with a standard deviation sampled uniformly between $\sqrt{2}$ and 2. Nodes without parents follow a Gaussian distribution with standard deviation sampled uniformly between $5\sqrt{2}$ and 10. The functions $f_{jk}$ are sampled from a Gaussian process with Gaussian kernel
and bandwidth one.
\begin{figure}[ht]
    \centering
    \includegraphics[width=0.75\linewidth]{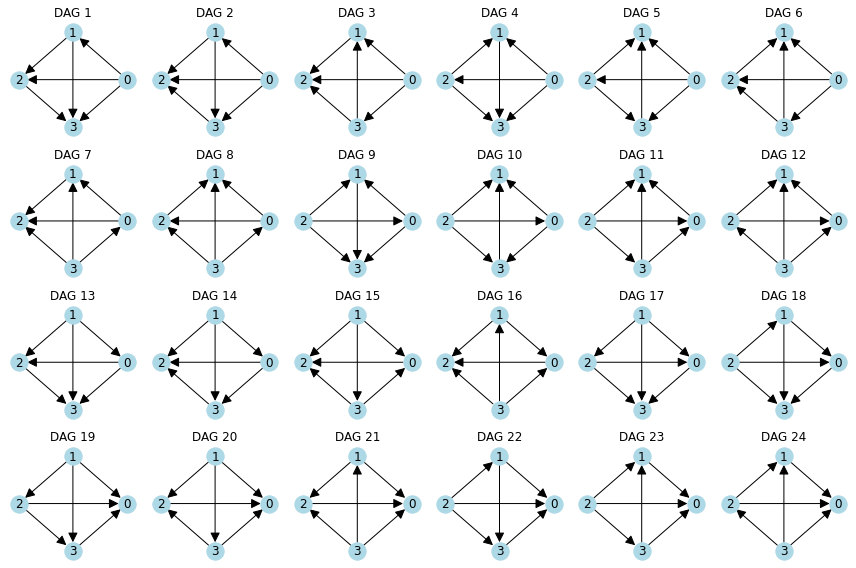}
    \caption{\textbf{Fully connected DAG of 4 nodes.} We use DAG 1 as the ground truth causal structure to simulate the data.}
    \label{fig: DAG4}
\end{figure}

In this example we simulate the data from DAG 1 and compare all the possible fully connected DAGs shown in Figure~\ref{fig: DAG4}. The procedure of the causal discovery is outlined as the following: (i) Run the generalised additive noise model for each DAG and get the residuals. (ii) Check the joint independence between the $d$ residuals. (iii) Report the DAG with the largest p-value.
\subsection{Real-World Financial Data}
The CPU time for computing high-order interaction percentages took ~8 hours without any parallelisation on a 2015 iMac with 4 GHz Quad-Core Intel Core i7 processor and 32 GB 1867 MHz DDR3 memory. We estimate the time taken for the permutation-based methods would have been at least 2 weeks. This highlights the scalability of our proposed method.
\begin{figure}[h]
    \centering
    \includegraphics[width=0.3\linewidth]{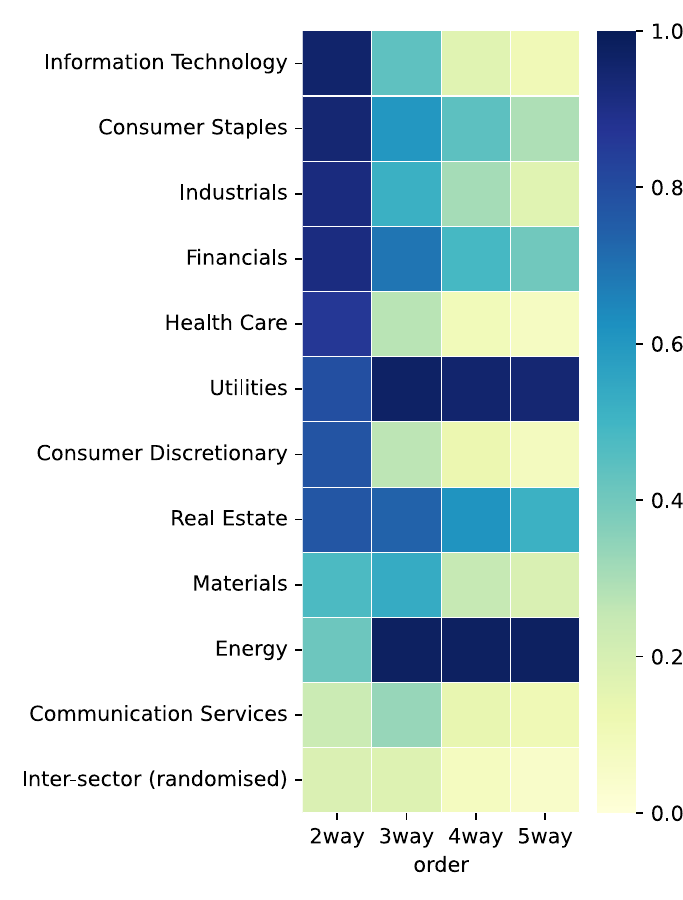}
    \caption{\textbf{High-order interactions percentages in between stocks in S\&P500.}  Heatmap of the percentages shown in Figure~\ref{fig: sp500} summarising the differences across sectors in their high-order interaction structure, also relative to randomised sets (last row). }
    \label{fig: sp500_heatmap}
\end{figure}
\end{document}